\documentclass{aa}

\usepackage{graphics}

\begin{document} 

\title{The Westerbork HI Survey of Spiral and Irregular
  Galaxies\\ I. HI Imaging of Late-type Dwarf Galaxies}

\author{R.A. Swaters\inst{1,2,3}\and T.S. van Albada\inst{1}\and J.M.
  van der Hulst\inst{1}\and R. Sancisi\inst{4,1}}

\institute{Kapteyn Astronomical Institute, P.O.~Box 800, 9700 AV
  Groningen, The Netherlands\and Dept. of Physics and Astronomy
  Johns Hopkins University, 3400 N. Charles Str., Baltimore, MD
  21218, U.S.A. \and Space Telescope Science Institute, 3700 San
  Martin Drive, Baltimore, MD 21218, U.S.A.  \and Osservatorio
  Astronomico di Bologna, via Ranzani 1, 40127 Bologna, Italy}

%\offprints{R.A. Swaters, \email{swaters@dtm.ciw.edu}}

\date{Received date; accepted date}

\def\HI{\ion{H}{i}}
\def\HII{H\kern1pt{\small II}}
\def\figHI{\ion{H}{i}}
\def\kms{km s$^{-1}$}
\def\etal{{et al.}}
\def\varv{v}
\def\clean{{\sc clean}}
\def\msun{\ifmmode\hbox{~M}_\odot\else$\hbox{M}_\odot$\fi}
\def\pc2{pc$^{-2}$}
\def\notes#1{\noindent{\sl #1\/}}
\font\ssymb=pzdr at 7pt
\def\star{{\ssymb H}}

\abstract{
Neutral hydrogen observations with the Westerbork Synthesis Radio
Telescope are presented for a sample of 73 late-type dwarf galaxies.
These observations are part of the WHISP project (Westerbork \HI\ 
Survey of Spiral and Irregular Galaxies).  Here we present \HI\ 
maps, velocity fields, global profiles and radial surface density
profiles of \HI, as well as \HI\ masses, \HI\ radii and line widths.
For the late-type galaxies in our sample, we find that the ratio of
\HI\ extent to optical diameter, defined as 6.4 disk scale lengths,
is on average $1.8\pm 0.8$, similar to that seen in spiral galaxies.
Most of the dwarf galaxies in this sample are rich in \HI\, with a
typical $M_{\HI}/L_B$ of 1.5.  The relative \HI\ content
$M_\mathrm{\HI}/L_R$ increases towards fainter absolute magnitudes
and towards fainter surface brightnesses. Dwarf galaxies with lower
average \HI\ column densities also have lower average optical
surface brightnesses.  We find that lopsidedness is as common among
dwarf galaxies as it is in spiral galaxies. About half of the dwarf
galaxies in our sample have asymmetric global profiles, a third has
a lopsided \HI\ distribution, and about half shows signs of
kinematic lopsidedness.
}

\titlerunning{HI Imaging of Late-type Dwarf Galaxies}
\authorrunning{R.A. Swaters et al.}   

\maketitle

\section{Introduction}
\label{theintro}

Over the past decades there have been numerous surveys of the \HI\ 
properties of galaxies using single dish telescopes. These have been
useful for determining the global \HI\ properties of galaxies and
for finding relations between \HI\ content, morphological type and other
global properties.  They have been particularly useful for obtaining
accurate redshifts and line widths for many galaxies.  However, due to
the large beam size, imaging galaxies in \HI\ was restricted to the
largest galaxies.

With the advent of synthesis radio telescopes \HI\ imaging became
routinely possible. Their high spatial resolution inspired many
studies of the detailed distribution and kinematics of \HI\ in
galaxies.  However, most of these studies focused on one or a few
galaxies, and only a few studies aimed at obtaining a large sample of
field galaxies.  Bosma (1978, 1981a,b) and Wevers (1984) both
investigated a sample of about 20 galaxies in \HI, but these studies
focused on large and \HI\ bright galaxies.  Broeils \& van Woerden
(1994) and Rhee \& van Albada (1996) investigated larger samples of
about 50 galaxies and over a wider range of galaxy properties, but
both studies used short observation with the Westerbork Synthesis
Radio Telescope (WSRT) so that essentially only one-dimensional data
are available.  In addition to these studies, there have also been
studies of the \HI\ distribution in cluster galaxies.  Warmels (1988a)
and Cayatte \etal\ (1990) studied the Virgo Cluster, Verheijen \&
Sancisi (2001, hereafter VS) the Ursa Major cluster.

To date, little effort has been put into making a large, homogeneous
\HI\ imaging survey of the local galaxy population.  Clearly, such a
survey is important for a systematic study of the \HI\ component,
including its kinematical properties and their relation with optical
properties.  For this reason, the Westerbork \HI\ Survey of Spiral and
Irregular Galaxies (WHISP) with the Westerbork Synthesis Radio
Telescope (WSRT) was started. The WHISP survey aims at mapping about
500 spiral and irregular galaxies in \HI.  There are two main aspects
to the WHISP project.  First, a systematic study of the \HI\ component
in itself, covering topics such as the distribution and kinematics of
\HI\ in and around galaxies, the frequency of lopsidedness and tidal
features in \HI, and the variation of these \HI\ properties with the
luminous properties and morphological types.  Second, the provision of
data for other studies, not directly related to the \HI\ component.
For example, the kinematic data are of great importance for
studies of rotation curves and dark matter properties, the frequency
of warps and the understanding of spiral structure.  For statistically
meaningful results, and for studies of variations with morphological
type and luminosity, many hundreds of galaxies need to be observed.
The WHISP project will provide the \HI\ maps and velocity fields
necessary for this work.  These data will be useful for other studies
of individual galaxies as well. In addition to the \HI\ data, for all
the galaxies in the WHISP sample $R$-band CCD photometry has also been
obtained (Swaters \& Balcells 2000, hereafter Paper~II). These will be
used to obtain photometric parameters such as integrated magnitudes,
surface brightnesses scale lengths.

This paper focuses on the WHISP \HI\ data for late-type dwarf
galaxies.  These galaxies in particular have been underrepresented in
previous \HI\ synthesis surveys.  Nonetheless, as has already been
found, these galaxies have interesting properties.  Single dish
observations have shown that they are generally rich in \HI, often
with $M_\mathrm{\HI}/L_B$ ratios larger than unity (e.g., Roberts \&
Haynes 1994). The largest \HI\ disks, relative to the optical extent,
are found in late-type dwarf galaxies, such as NGC~2915 (Meurer \etal\ 
1996), DDO~154 (Carignan \& Purton 1998) and NGC~4449 (Hunter \etal\ 
1998).  Rotation curve studies show that these galaxies may have even
larger dark matter fractions than ordinary spirals (e.g., Carignan \&
Beaulieu 1989; Broeils 1992b; Swaters 1999; C{\^o}t{\'e} \etal\
2000).

In this paper we present \HI\ observations for a sample of 73
late-type dwarf galaxies and we compare their \HI\ properties with
their luminous properties. Sect.~\ref{thesample} describes the
sample selection and the observations. Sect.~\ref{thereduction}
describes the reduction of the data and the derivation of the \HI\ 
maps and velocity fields, the radial and global profiles, and global
properties such as the \HI\ mass, line widths and \HI\ diameters.  In
section~\ref{theprops} the \HI\ properties of these galaxies are
compared to the optical properties, and to the \HI\ and optical
properties of bright spiral galaxies. Sect.~\ref{themorphology}
discusses the occurrence of asymmetries in late-type dwarf galaxies.
In Sect.~\ref{thesummary} the conclusions are presented.

To facilitate the reading of this paper, all long tables have been
put into Appendix~A. An atlas of the \HI\ observations is
presented in Appendix~B.

\begin{figure}
\resizebox{\hsize}{!}{\includegraphics{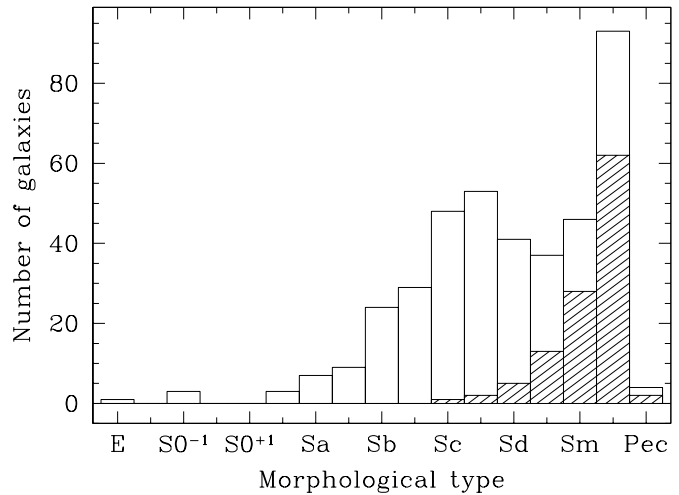}}
\caption{Histogram of Hubble types in the primary WHISP sample. The
  shaded area gives the distribution over Hubble types of the sample
  presented in this paper.}
\label{fighisttypes}
\end{figure}

\section{The sample}
\label{thesample}

The galaxies for the WHISP survey have so far been selected from the
Uppsala General Catalogue of Galaxies (UGC, Nilson 1973), taking
observability with the WSRT into account.  The typical resolution of
the WSRT is $14''\times 14''/\sin\delta$. To make sure that the
galaxies will be sufficiently resolved by the WSRT beam, we first
selected galaxies with blue diameters larger than $1.5'$ and
declinations north of $20^\circ$.  This resulted in a sample of 3148
galaxies. For 1560 of these \HI\ fluxes are catalogued in the
Third Reference Catalogue of Bright Galaxies (hereafter RC3, de
Vaucouleurs et al.\ 1991).  Next, the sample was restricted further to
galaxies with known \HI\ flux densities larger than 100 mJy, where the
\HI\ flux density is the ratio of the integrated \HI\ flux and profile
width at the 50\% level as listed in the RC3.  This resulted in our
primary WHISP sample of 409 galaxies.  The distribution over
morphological types of this sample is shown in
Fig.~\ref{fighisttypes}.  Almost half of the galaxies in the sample
selected as described above are late-type dwarf galaxies.

From the primary sample we selected late-type dwarf galaxies, i.e.,
all galaxies with Hubble types later than Sd, supplemented with
galaxies of earlier Hubble type, but with absolute magnitudes fainter
than $M_B=-17$. Furthermore, we only selected galaxies with \HI\ flux
densities larger than 200 mJy and with Galactic latitudes
$|b|>10^\circ$.  This resulted in a sample of 109 galaxies. To this
sample we added galaxies that met all selection criteria of the
primary WHISP sample, except the diameter criterion. This additional
sample consists of 4 galaxies, and these were added to the sample
presented in this paper because of the possible uncertainties in using
the optical diameter $D_{25}$ as an indicator of the \HI\ diameter
(see Sect.~4.1). The combined sample constructed in this way
consists of 113 galaxies.  The distribution over morphological types
of the sample is shown in Fig.~\ref{fighisttypes} as the shaded area.
Because the selection is largely based on morphological type, a few
galaxies have been included in the sample that have been classified as
irregulars, but that proved to be large interacting galaxies rather
than dwarf galaxies.  Their data are presented here as well.

Out of the sample thus selected, 73 have been observed (these are
listed in Table~\ref{tabsample}).  For most of the 40 other galaxies
\HI\ data have been obtained by others, although some of these data
are as yet unpublished.  These 40 galaxies will not be discussed here.

The adopted distances for the galaxies in our sample are listed in
Table~\ref{tabsample}. If available, distances from Cepheids,
brightest stars or group membership were used. In case none of these
were available, the distance was derived from the systemic velocity,
adopting a Hubble constant of $H_0=75$ km s$^{-1}$ Mpc$^{-1}$, and
correcting for the Virgocentric inflow following the prescription
given in Kraan-Korteweg (1986). A more detailed discussion of the
adopted distances and the distance uncertainties is given in Paper~II.

\section{Observations and reduction}
\label{thereduction}

The \HI\ data presented in this paper were obtained with the WSRT
between 1992 and 1996. Most galaxies were observed in a single 12 hour
synthesis observation. Large galaxies were observed longer to increase
the radius of the first grating ring. The typical full resolution is
$14''\times 14''/\sin\delta$.  The observations were done with a uniform
velocity taper and a channel separation of 2 or 4 \kms.  The
observational setup for each galaxy and the reduction steps are
detailed in the WHISP webpages, http://www.astro.rug.nl/$\sim$whisp.
Here, we present only a summary of the reduction steps. Unless
otherwise stated, the reduction procedure outlined below is the WHISP
pipeline reduction as detailed on the WHISP webpages.

The raw UV data were calibrated and flagged interactively, using the
NEWSTAR software developed at NFRA at Dwingeloo. Next, the UV data
were Fourier transformed to the map plane. Three sets of maps were
produced for each galaxy, at the full resolution, at $30''\!\times
30''$ and at $60''\!\times 60''$. For each data cube five antenna
patterns were calculated, spread evenly through the data cube.

After the maps were constructed, the continuum was subtracted by
fitting and subtracting a first order polynomial to the channel maps
without \HI\ line emission. Next, the maps were {\clean}ed in the map
plane in two steps. First, the areas with \HI\ emission were
identified in the $60''$ Hanning-smoothed data and masks indicating
the areas with emission were constructed by hand. The data were then
{\clean}ed, using the masks to define the search areas.  Next, the
\clean\ components were restored and these data were used to refine
the masks.  These masks were then used to define the search areas for
all three resolutions.  Each data cube was {\clean}ed with antenna
patterns at the appropriate resolution down to $0.5\sigma$.

\subsection{Global profiles, line widths and \HI\ masses}

The global profiles were constructed from the flux in the {\clean}
components obtained from the $60''$ Hanning-smoothed data cubes, after
correction for primary beam attenuation. Because the maps were
{\clean}ed down to $0.5\sigma$, there is still a small amount of
residual flux in the map. In principle, this residual flux may be
added to the flux in \clean\ components to give the total flux. There
is a pitfall, however. The residual flux is given by the sum over the
search area divided by the sum over the antenna pattern.  The sum over
the antenna pattern varies with the size of the box and may even
change sign, possibly resulting in inaccurate values for the residual
flux.  Therefore, only the flux in the \clean\ components was used to
calculate the global profile. On average, the total flux including the
residual flux is 2\% larger than the flux in the \clean\ components
only. Only for 10\% of the galaxies, the differences between the two
fluxes is larger than 5\%, but it is always smaller than 8\%.  The
global profiles for all galaxies are shown in Appendix~B.

The \HI\ flux integral for each galaxy has been obtained by adding the
fluxes in the \clean\ components, correcting for primary beam
attenuation. From this, the \HI\ mass was calculated with the standard
formula
\begin{equation}
\mathrm{M}_\mathrm{\HI}=236\cdot\mathrm{D}^2\cdot\mathrm{S},
\label{himass}
\end{equation}
where $\mathrm{M}_\mathrm{\HI}$ is the \HI\ mass in
$\mathrm{M}_\odot$, D the distance in Mpc and S the \HI\ flux integral
in mJy \kms. The flux integrals and \HI\ masses are listed in
Table~\ref{tabprops}.

Flux measurements with a synthesis array may underestimate the total
flux because of the missing zero and short spacings. The shortest
spacings for the present observations were 36 or 72 m, therefore the
observations are less sensitive to structures more extended than 5 or
10 arcmin. Because most of the galaxies in our sample are smaller than
this limit, the fluxes are in general expected to be well determined.
VS found that the WSRT \HI\ fluxes of the galaxies in his sample are
in excellent agreement with those obtained from single dish
observations (the galaxies in his sample are all smaller than $10'$).
Comparison of the flux densities derived from the observations
presented in this paper to the ones derived from the RC3 data, as used
to select our sample, confirms that we are not systematically missing
flux. The derived flux densities generally agree, although with a
large scatter of about 15\%. For galaxies with \HI\ diameters larger
than $400''$ we seem to miss up to about 10\% to 15\% of the single dish
flux.

From the global profiles the line widths at the 20\% and 50\% level
were derived. For double-horn profiles, the peaks on both sides were
used separately to calculate the 20\% or 50\% levels.  In other cases,
the overall peaks of the profiles were used.  The line-width
$W_{20}^\mathrm{obs}$ ($W_{50}^\mathrm{obs}$) was defined as the
difference between the velocities at the 20\% (50\%) levels on both
sides of the global profile.  The systemic velocity
$\varv_\mathrm{sys}$ was taken to be the average of the midpoints
between the profile edges at the 20\% and 50\% level.

The line widths have been corrected for instrumental broadening using
expressions given by VS:
\begin{equation}
\begin{array}{l}
W_{20}= W_{20}^\mathrm{obs}-35.8\left[\sqrt{1+\left(R\over{23.5}\right)^2}-1\right] \\
\\
W_{50}= W_{50}^\mathrm{obs}-23.5\left[\sqrt{1+\left(R\over{23.5}\right)^2}-1\right],
\end{array}\label{eqninstr}
\end{equation}
where $R$ is the instrumental velocity resolution in \kms. Corrections
for broadening due to random motions of the \HI\ gas have been
calculated with expressions given by Tully \& Fouqu\'e (1985):
\begin{equation}
\begin{array}{rl}
W^2_{R,l} \; = \; W^2_l \kern-7pt & + \; W^2_{t,l}\;\left[1 - 
2\;e^{-\left(\frac{W_l}{W_{c,l}}\right)^2}\right]\\
& \\
& - \; 2\;W_l\;W_{t,l}\left[1 -
e^{-\left(\frac{W_l}{W_{c,l}}\right)^2}\right],
\end{array}\label{eqnrandom}
\end{equation}
where the subscript $l$ refers to the line width at the $l=20$\% or
the $l=50$\% level, and $W_{t,l}$ is a term that represents random
motions. This formula gives a quadratic subtraction if $W_l<W_{c,l}$,
and a linear subtraction if $W_l>W_{c,l}$ We have adopted
$W_{c,20}=120$ \kms, $W_{c,50}=100$ \kms, $W_{t,20}=32$ \kms\ and
$W_{t,50}=15$ \kms\ (VS).

\subsection{Integrated \HI\ maps}

The integrated \HI\ maps were constructed by adding the signal from
the {\clean}ed channel maps within the areas defined by the \clean\ 
masks. Integrated \HI\ maps were constructed at the full resolution,
and at $30''\!\times 30''$ and $60''\!\times 60''$, using the same
masks. These \HI\ maps were corrected for primary beam attenuation.

The \HI\ maps at $30''$ resolution are shown in Appendix~B.  All plots
have been displayed on a common gray scale. Note that the noise in an
integrated \HI\ map is not constant over the map, because at each
position data were added from a different number of channel maps due to
the masking used. For uniformly tapered data sets, the noise is given
by:
\begin{equation}
\sigma_{tot}=n_l^{0.5}\sigma_c\sqrt{(n_c+n_l)/(n_c-1)},
\end{equation}
where $\sigma_{tot}$ is the noise in the integrated \HI\ map and
$\sigma_c$ the noise in the fit to the line-free channels
used to subtract the continuum.  The number of channels contributing
to the integrated \HI\ maps is given by $n_l$, and the number of
line-free continuum channels is denoted by $n_c$. With this equation a
noise map was calculated, which was then used to construct a
signal-to-noise map.  Using the signal-to-noise map, we selected from
the integrated \HI\ map all pixels with a signal-to-noise ratio
between 2.75 and 3.25. The average value of these points was then used
to define the $3\sigma$ contour indicated by the thick line in the
integrated \HI\ maps in the figures in Appendix~B.

\begin{figure}
\resizebox{\hsize}{!}{\includegraphics{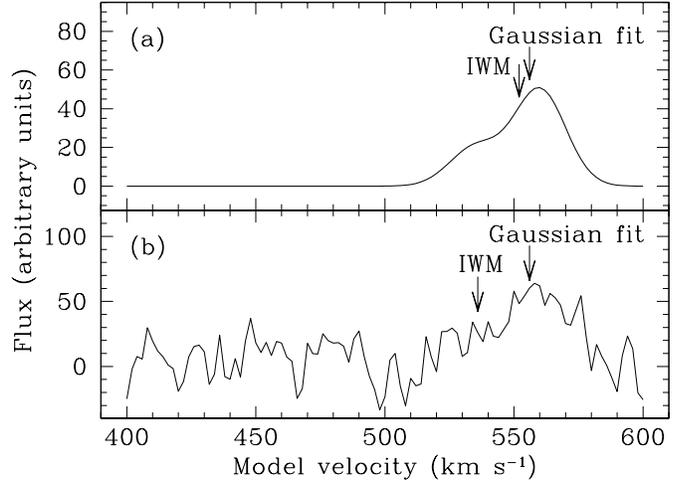}}

\caption{ ({\bf a}) The intensity weighted mean (IWM) and a Gaussian
  fit to a skewed model profile. Both the Gaussian fit and the IWM
  fail to recover the peak velocity, which represents the rotation
  velocity. The Gaussian fit is closer to the input rotation velocity.
  ({\bf b}) The same profile as in panel a, but with a SNR of 3. The
  Gaussian fit velocity is hardly affected, but the IWM velocity is
  strongly influenced by the noise. For details see text. }
\label{figiwmvsgau}
\end{figure}

\subsection{Velocity fields}

The velocity fields were derived in two steps. First, as part of the
standard WHISP reduction, a moment analysis was done, which yielded
the intensity weighted mean (IWM) velocity field.  It is well known
that the IWM velocities may suffer from systematic errors.  For
example, if a profile is skewed (e.g., as a result of beam smearing,
high inclination or a thick \HI\ disk), the IWM will yield a value
offset from the peak towards the skewed side, as demonstrated in
Fig.~\ref{figiwmvsgau}a.  A second possible source of systematic error
in the IWM velocities is that profiles with low signal-to-noise ratios
will be biased to the midpoint of the velocity range of the profile
(see Fig.~\ref{figiwmvsgau}b).  For a more detailed discussion of the
systematic effects that may be introduced if IWM velocities are used,
see Swaters (1999). 

A better velocity determination is obtained by fitting a single Gauss
to the velocity profile, which is less sensitive to asymmetries and
noise (see Fig.~\ref{figiwmvsgau}). We therefore determined the
velocity fields by fitting a single Gaussian to the line profiles at
each position in the Hanning smoothed $30''$ resolution data. Initial
estimates for the fits were obtained from the moment analysis. Only
those fits were used that had amplitudes higher than $3\sigma$.  The
velocity fields derived in this way are shown in Appendix~B.

\subsection{Radial \HI\ surface density profiles and \HI\ diameters}

The total \HI\ maps have been used to derive radial surface density
profiles of \HI. Two different methods were used, depending on
inclination and angular size.  For galaxies with inclinations smaller
than about $75^\circ$ that are well resolved, the data were
azimuthally averaged in concentric ellipses. The orientation
parameters used for this are the same as those for the rotation curve
analysis (see Swaters 1999). Mostly, these are the same as the optical
orientation parameters. The azimuthal averaging was done for the
approaching and the receding sides separately. Pixels without measured
signal in the total \HI\ map were excluded.

Azimuthal averaging following the above procedure does not produce
reliable results for highly inclined galaxies or galaxies that are
poorly resolved. For these galaxies the \HI\ radial surface density
profiles were derived following the method described by Warmels
(1988b). First the total \HI\ maps were integrated parallel to the
minor axis, resulting in \HI\ strip integrals. To get the \HI\ surface
density profiles, the iterative deconvolution scheme described by Lucy
(1974) was applied to the \HI\ strip integrals, with the assumption
that the \HI\ distribution is axisymmetric.  In short, this method
works as follows. An input estimate of the radial profile is converted
to a strip integral and smoothed to the resolution of the
observations. Next, this strip integral is compared to the observed
one, and the input profile is adjusted, and used as input for the next
cycle, until a convergence criterion is met. For a detailed
description of the procedure, see Warmels (1988b).

\begin{figure}
\resizebox{\hsize}{!}{\includegraphics{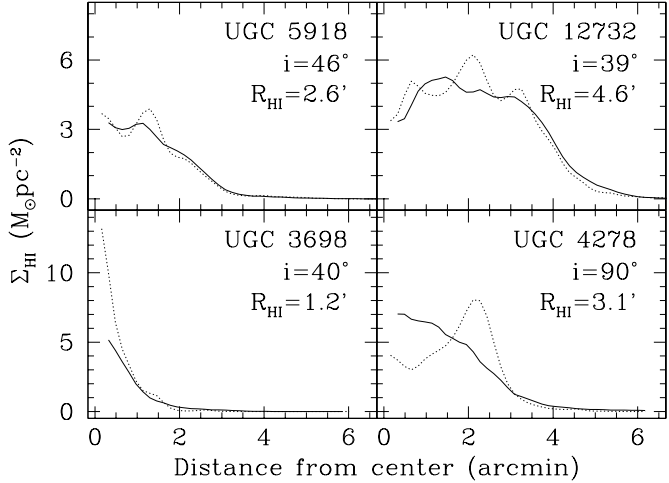}}
\caption{Comparison between the radial \figHI\ surface density profiles
  derived from averaging in ellipses (full line) and using the Lucy
  deconvolution scheme (dotted line).}
\label{figradhi}
\end{figure}

In Fig.~\ref{figradhi} a comparison is shown between the profiles
derived from averaging in ellipses and those based on the Lucy
deconvolution scheme. For most galaxies that are not highly inclined,
the two profiles agree well, as is seen in Fig.~\ref{figradhi} for
\object{UGC~5918} and \object{UGC~12732}. However, for small and
highly inclined galaxies the profiles differ, as expected. For small
galaxies, illustrated by \object{UGC~3698} in Fig.~\ref{figradhi}, the
central parts are not resolved, and the Lucy deconvolution restores
some of the flux to the central parts. A known problem with this
method is that it may restore too much flux to the center. To avoid
this, the iteration process to determine the radial profile was
stopped after 10 iterations, or at an earlier stage if the derived
radial profile matched the observed strip integral at the 95\%
confidence level, following Warmels (1988b).  For edge-on galaxies,
averaging over ellipses obviously gives an incorrect result. If the
galaxies are optically thin in \HI, the Lucy method will give the
correct radial \HI\ surface density profile.

The surface density profiles have been used to determine the \HI\ 
radii, $R_\mathrm{\HI}$, defined as the radius where the \HI\ surface
density corrected to face-on reaches a density of $1\msun/\mathrm{
  pc}^{-2}$. These radii are given in Table~\ref{tabprops}.

Most \HI\ surface density profiles are close to exponential in their
outer parts. This can be clearly seen in Fig.~\ref{figexpfits}, where
the profiles have been plotted on a logarithmic scale and the dotted
lines give the exponential fits made to the profile.  The derived \HI\ 
scale lengths are given in Table~\ref{tabprops}.

Finally, the average \HI\ surface density within 3.2 optical disk
scale lengths ($\langle\Sigma_\mathrm{\HI}\rangle_{3.2h}$, where $h$
is the optical disk scale length) was
determined, with the optical scale lengths as determined in Paper~II.
The derived average \HI\ surface densities are listed in
Table~\ref{tabprops}.  Use of the optical scale length of the disk has
certain advantages over the isophotal radius.  The number of optical
disk scale lengths within an isophotal radius depends on the central
disk surface brightness.  The lower the surface brightness, the
smaller the number of optical scale lengths within the isophotal
radius. For galaxies with very low surface brightnesses, the central
surface brightness may even be fainter than the chosen isophotal value
and the isophotal diameter would therefore not be defined. The
determination of the optical disk scale length, on the other hand, is
independent of surface brightness.

To keep our definition of galaxy diameter compatible with the often
used isophotal diameters, we defined the galaxy diameter as $6.4h$.
This value was chosen because most spiral galaxies for which the
isophotal diameter $D_{25}$ has been measured, have an approximately
constant central disk surface brightness of 21.65 $B$-mag
arcsec$^{-2}$ (Freeman 1970), and for these galaxies the 25 mag
arcsec$^{-2}$ isophote is reached at $3.2h$.  Hence, the definition of
$6.4h$ as a measure of the galaxy diameter is similar to $D_{25}$ for
bright spiral galaxies, but it gives a consistent definition of the
galaxy diameter irrespective of surface brightness.

\begin{figure*}
\resizebox{0.875\textwidth}{!}{\includegraphics{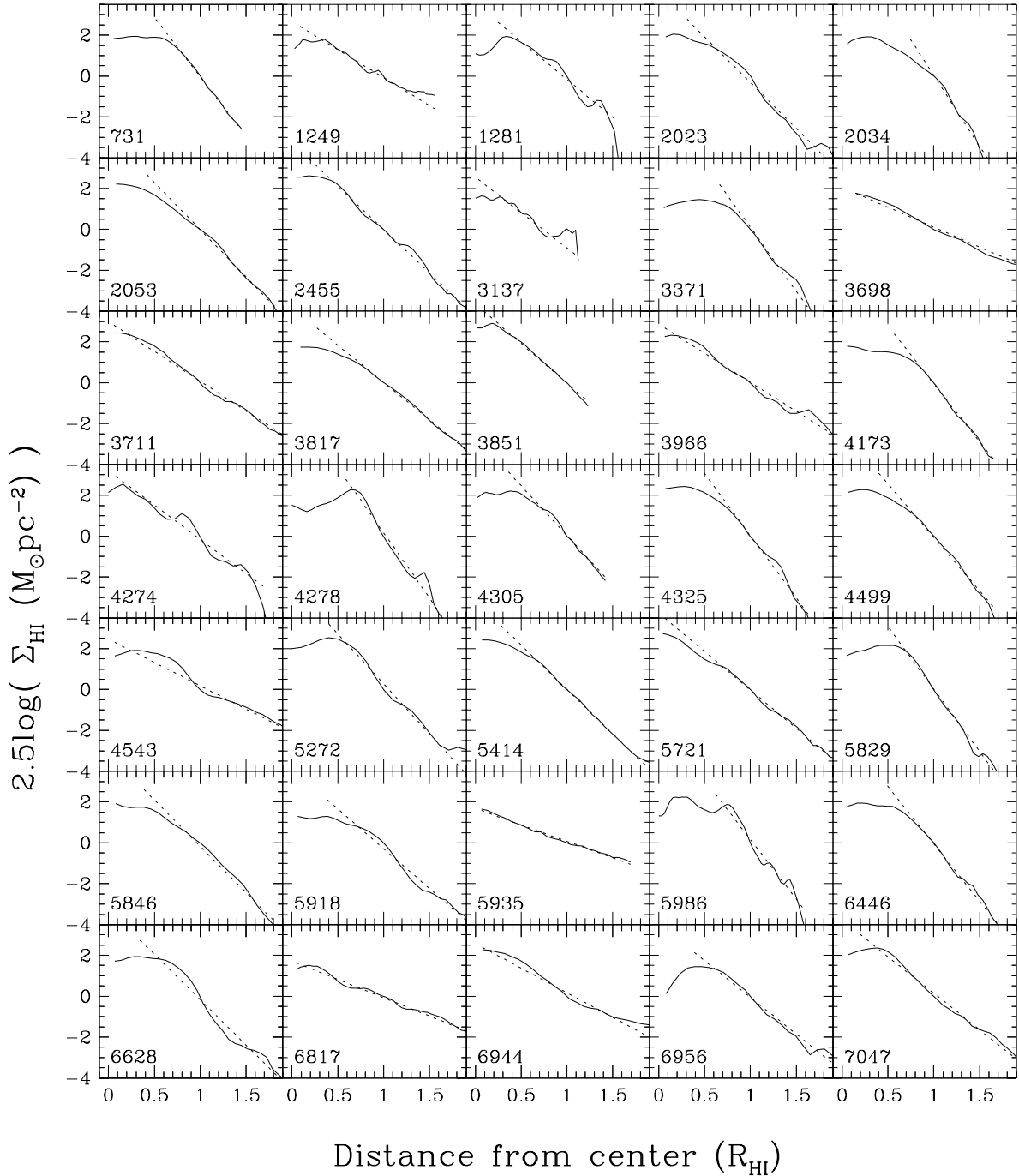}}
\caption{ Radial \figHI\ surface density profiles (full lines) plotted
  on a `magnitude scale', i.e., $2.5\log \Sigma_\mathrm{\HI}$, so that
  the \figHI\ scale lengths are defined in the same way as the optical
  disk scale lengths. The dotted lines represent the exponential fits to
  the outer parts of the profiles. The UGC numbers are given in the
  lower left corner of each panel.
}
\label{figexpfits}
\end{figure*}

\begin{figure*}
\resizebox{0.875\textwidth}{!}{\includegraphics{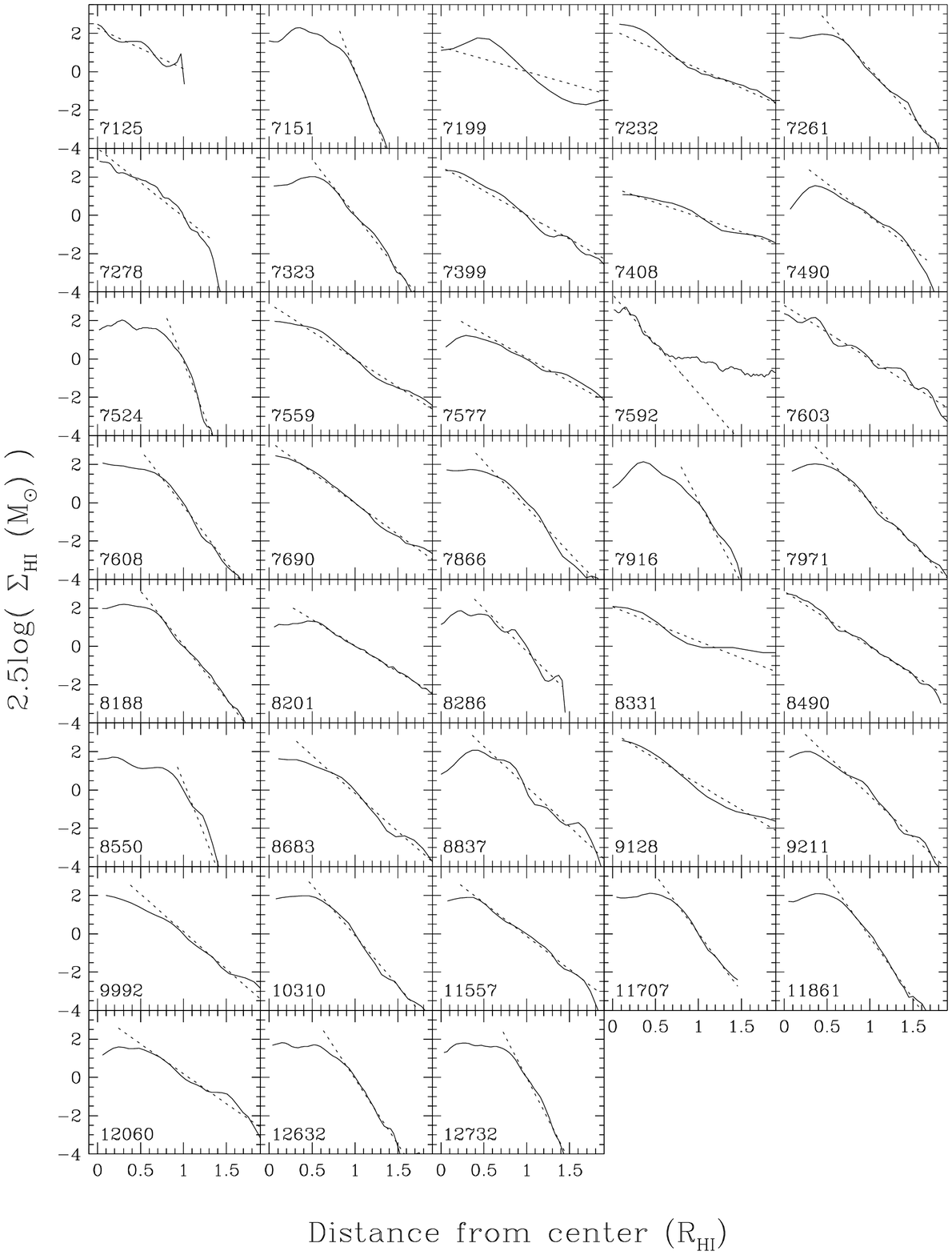}}
\addtocounter{figure}{-1}
\caption{continued}
\end{figure*}

The large sample of dwarf galaxies presented here allows a study of
the relations between the \HI\ properties of late-type dwarf galaxies
and their optical properties. It is important to find out whether the
relations found for bright spiral galaxies are also valid for the
dwarf galaxy regime.  As several studies on the relations between the
global properties of late-type dwarf galaxies, based on larger and
more complete samples, have already been published (e.g., Roberts \&
Haynes 1994; Hoffman \etal\ 1996), we will focus here on \HI\ radii
and surface densities.  We use the optical data presented in Paper~II.
Absolute magnitudes, extrapolated central disk surface brightnesses
and scale lengths are listed in Table~\ref{tabsample}.

\section{\HI\ properties}
\label{theprops}

\subsection{\HI\ diameters versus optical diameters}

\begin{figure}
\resizebox{\hsize}{!}{\includegraphics{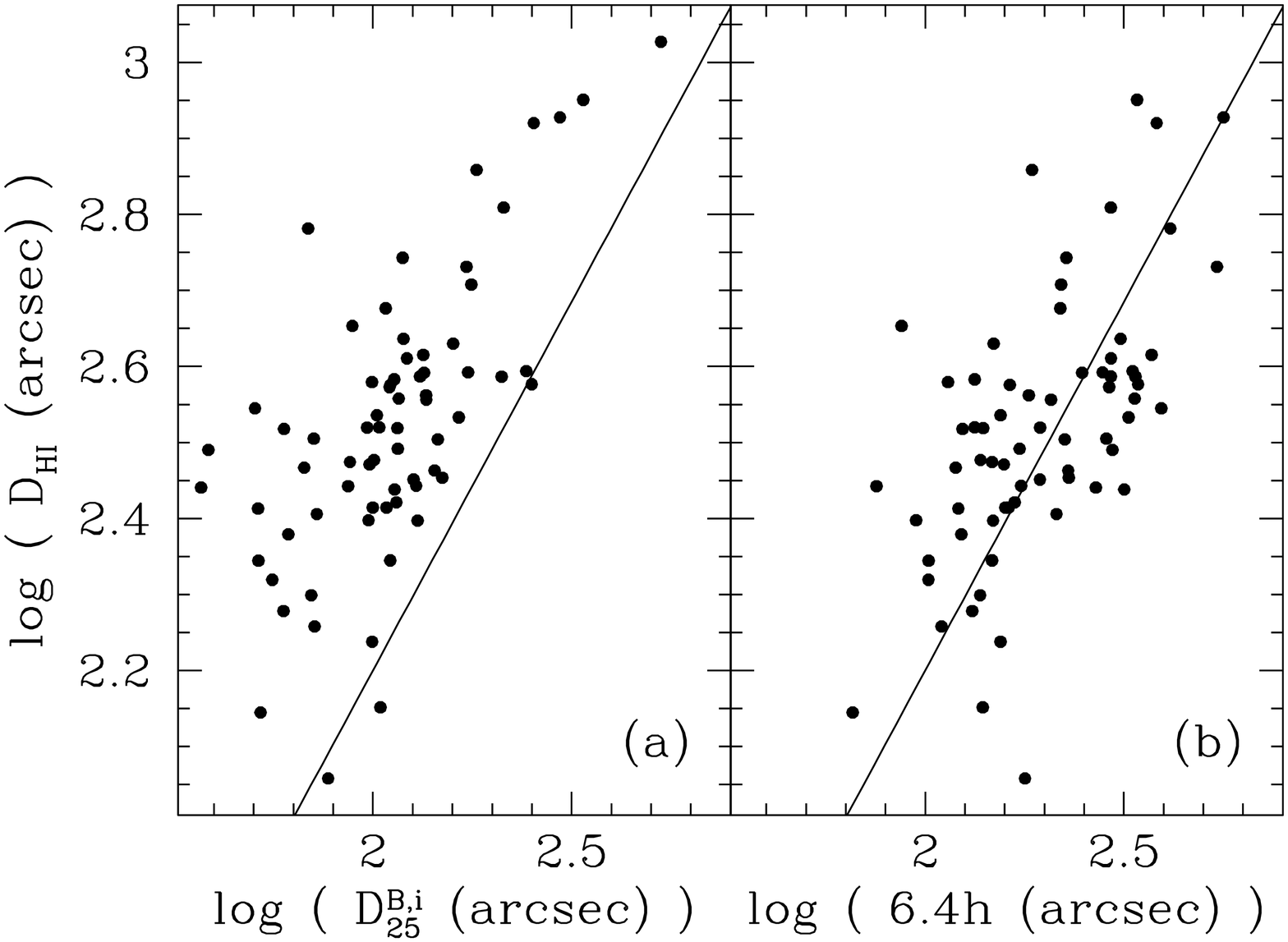}}
\vspace{-6mm}
\caption{The \figHI\ diameter $D_\mathrm{\HI}$ versus $D_{25}^{\,B,i}$ ({\bf
    a}) and $6.4h$ ({\bf b}). The full line in both panels represents
    the relation found between $D_\mathrm{\HI}$ and $D_{25}^{\,B,i}$ by
    Broeils \& Rhee (1997).}
\label{figradcomp}
\end{figure}

Broeils \& Rhee (1997, hereafter BR) have investigated the \HI\ 
properties of a sample of spiral galaxies mapped in \HI\ with the
WSRT. The sample they use is a combination of the samples of Broeils
\& van Woerden (1994) and Rhee \& van Albada (1996).  BR find a strong
correlation between the \HI\ diameter $D_\mathrm{\HI}$ (defined at a
\HI\ surface density of 1 \msun\pc2) and the absorption-corrected
optical diameter $D_{25}^{\,B,i}$, measured at the 25 $B$-mag
arcsec$^{-2}$. In Fig.~\ref{figradcomp}a a comparison is shown between
the optical and the \HI\ diameters for the late-type dwarf galaxies in
our sample. To make this comparison, our $R$-band diameters have been
transformed to $B$-band diameters assuming $D_{25}^{\,R,i}/
D_{25}^{\,B,i}=1.4$. This ratio was determined from the 46 galaxies
for which we have both $R$ and $B$-band data (see Paper~II). The full
lines in Fig.~\ref{figradcomp}a and b show the relation between
$D_\mathrm{\HI}$ and $D_{25}^{\,B,i}$ found by BR (we assumed that for
the BR sample $D_{25}^{\,B,i}$ equals $6.4h$).  Nearly all
galaxies in our sample have much larger \HI\ diameters than expected from
the relation found by BR. However, this is most likely due to the
choice of the definition of the optical diameter. As argued in
Sect.~\ref{thereduction}, the isophotal diameter $D_{25}^{\,B,i}$ is
not a suitable definition if galaxies with different central disk
surface brightnesses are compared. For galaxies with lower surface
brightnesses, only a smaller fraction of the disk is enclosed within
the isophotal diameter. Because most of the dwarfs in this sample have
low surface brightnesses, the use of the isophotal diameter
$D_{25}^{\,B,i}$ leads to small optical diameters, which could explain
the offset seen in Fig.~\ref{figradcomp}a.

\begin{figure}
\resizebox{\hsize}{!}{\includegraphics{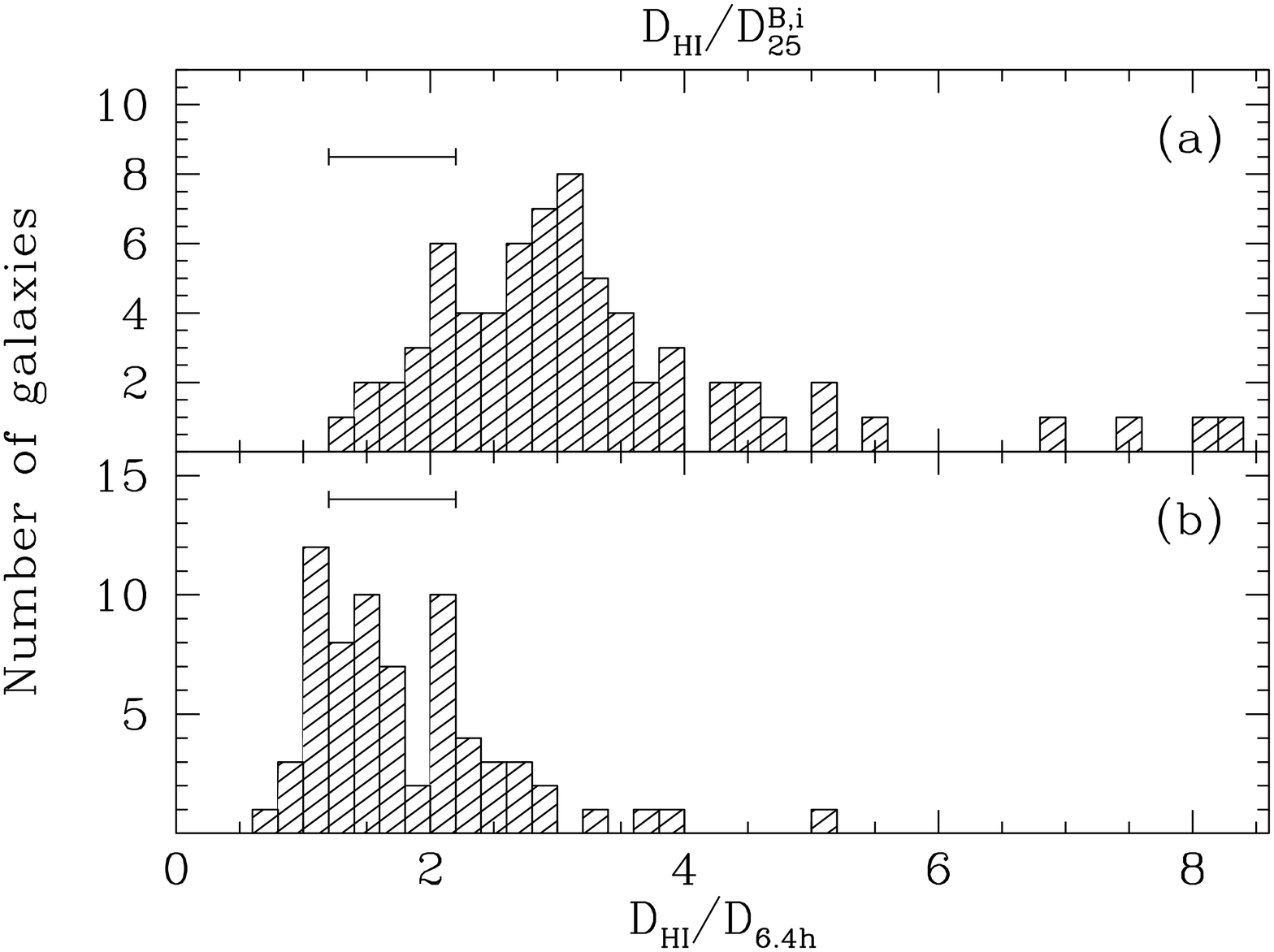}}
\vspace{-6mm}
\caption{Histogram of $D_\mathrm{\HI}/D_{25}^{\,B,i}$ (top panel) and of
  $D_\mathrm{\HI}/6.4h$ (bottom panel). The horizontal bar in both
  panels indicates the range of $D_\mathrm{\HI}/D_{25}^{\,B,i}$ for
  spiral galaxies, as found by Broeils \& Rhee (1997).}
\label{figradhist}
\end{figure}

In Fig.~\ref{figradcomp}b, the relation between optical diameter and
\HI\ diameter is shown, but now the optical diameter has been defined
as 6.4$h$, as argued above.  The scale lengths used in
Fig.~\ref{figradcomp}b are $R$-band scale lengths.  A comparison of
the $B$-band and $R$-band scale lengths for the 46 galaxies for which
both bands have been observed showed that on average the scale lengths
are similar in both bands (see Paper~II).  With the optical radius
defined as $6.4h$, the late-type dwarf galaxies follow the relation as
found by BR. But the relation for dwarf galaxies has a larger scatter
than found by BR for brighter spirals, indicating that the optical and
\HI\ diameters are less strongly coupled in the dwarf regime, or that
the optical and \HI\ diameters are less well defined.

The importance of the definition of optical radius is also seen
clearly in Fig.~\ref{figradhist}, where the distributions of
$D_\mathrm{\HI}/D_{25}^{\,B,i}$ and $D_\mathrm{\HI}/6.4h$ are shown.
The horizontal bar indicates the average found by BR of $1.7\pm 0.5$.
Expressed in units of $D_{25}^{\,B,i}$, the late-type dwarf galaxies
have extreme properties, with $D_\mathrm{\HI}/ D_{25}^{\,B,i}=3.3\pm
1.5$. In the more suitable units of $6.4h$, the gaseous extent of
dwarf galaxies is $D_\mathrm{\HI}/ D_{6.4h}=1.8\pm 0.8$. Summarizing,
we find that the average \HI\ extent of late-type dwarf galaxies
relative to the optical diameter is similar to that of bright spiral
galaxies, but with a larger spread.  In particular, it seems that the
most extended \HI\ disks, relative to the optical disks, are found
among dwarf galaxies (e.g., Meurer \etal\ 1996; Carignan \& Purton
1998; Hunter \etal\ 1998).

\begin{figure}
\resizebox{\hsize}{!}{\includegraphics{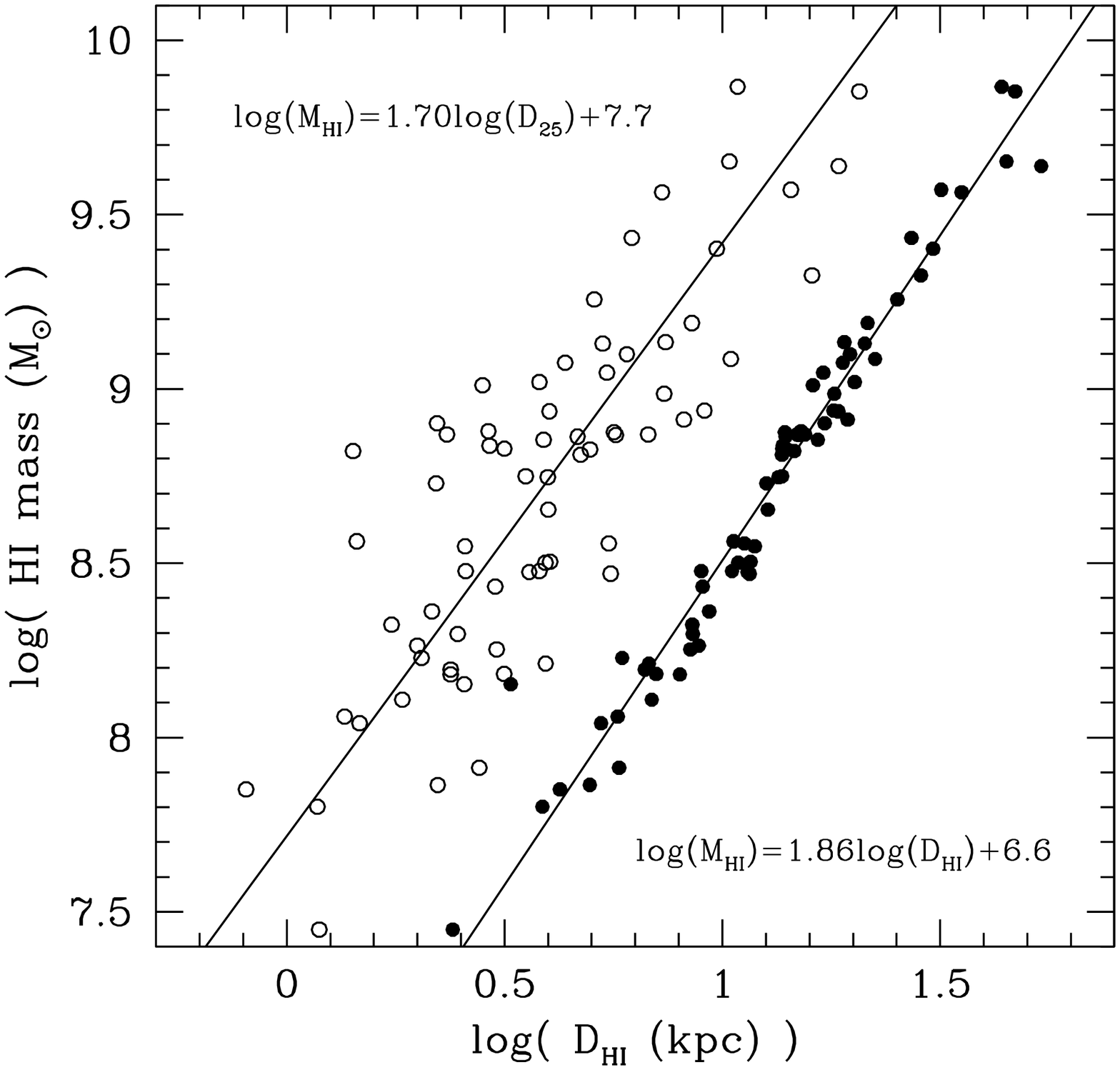}}
\vspace{-6mm}
\caption{ \figHI\ mass versus \figHI\ diameter (filled circles) and versus
  optical diameter ($6.4h$, open circles). The solid lines represent
  the fits to the data points. The resulting fit is printed next to
  the fitted line.
}
\label{figdiamhimass}
\end{figure}

As shown in previous studies (BR; VS), there is
a tight relation between \HI\ mass and \HI\ diameter. This relation
also holds for dwarf galaxies, as is shown in
Fig.~\ref{figdiamhimass}.  The small scatter about this relation
points to a small spread in mean \HI\ surface density for the dwarf
galaxies in our sample.  The \HI\ mass also correlates with the
optical diameter, shown in Fig.~\ref{figdiamhimass} as well, but with
a larger scatter.

\begin{figure}[ht]
\resizebox{\hsize}{!}{\includegraphics{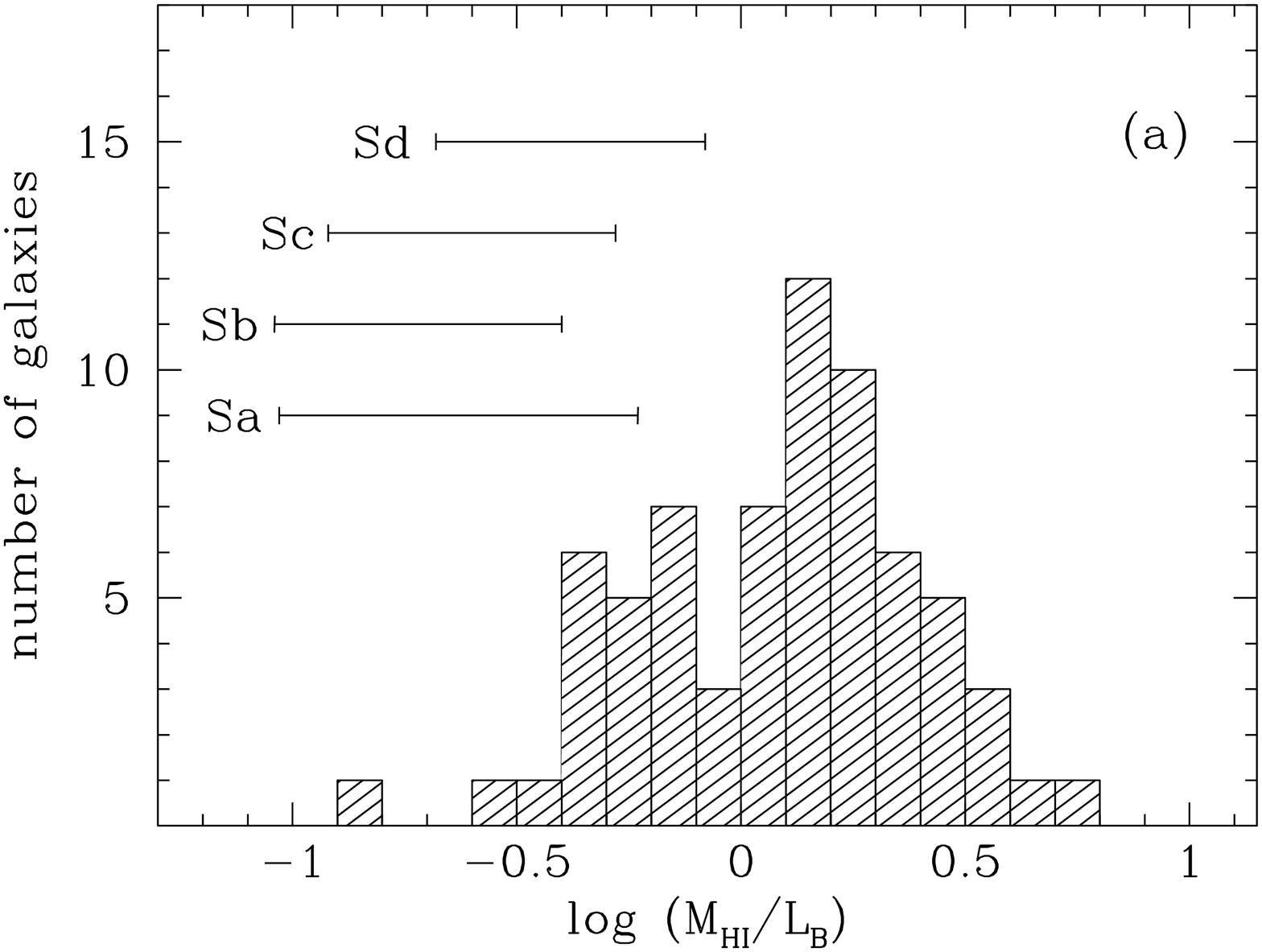}}
\caption{The distribution of $M_\mathrm{\HI}/\mathrm{L}_R$, the
  horizontal bars indicate the ranges of $M_\mathrm{\HI}/
  \mathrm{L}_R$ for other morphological types, as found by Broeils \&
  Rhee (1997).}
\label{fighistmhilb}
\vspace{10mm}
\resizebox{\hsize}{!}{\includegraphics{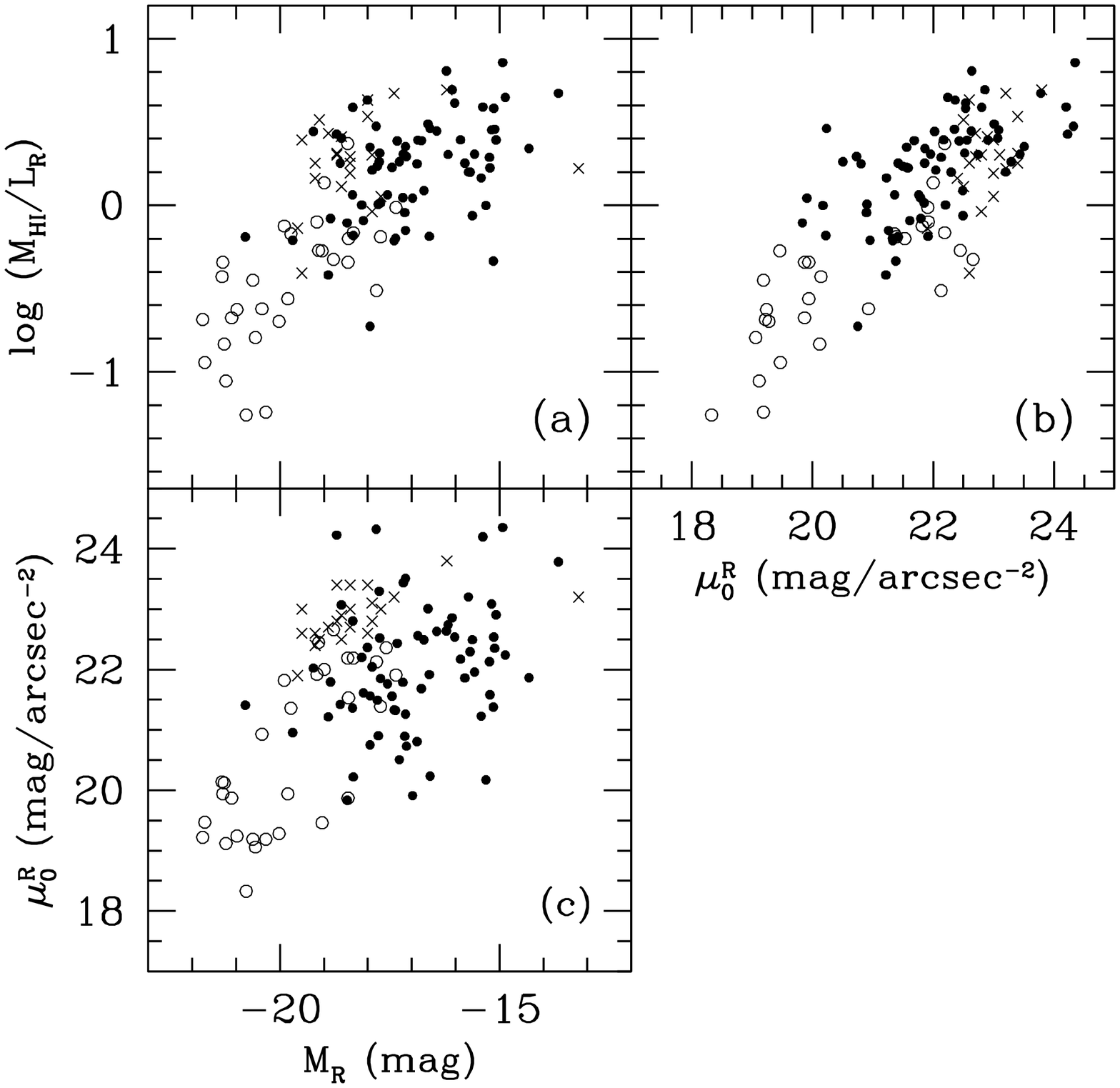}}
\vspace{-7mm}
\caption{ {\bf (a)} $M_\mathrm{\HI}/\mathrm{L}_R$
  against absolute $R$-band magnitude, filled circles: late-type
  dwarfs, open circles: Ursa Major galaxies (Verheijen \& Sancisi 1997), crosses:
  low surface brightness galaxies (de Blok \etal\ 1996). {\bf (b)}
  $M_\mathrm{\HI}/\mathrm{L}_R$ versus $R$-band surface
  brightness. {\bf (c)} $R$-band surface brightness versus $R$-band
  magnitude.  }
\label{figmhilrmu}
\end{figure}

\subsection{\HI\ mass versus luminosity}

Fig.~\ref{fighistmhilb} shows the distribution of $M_\mathrm{\HI}/
\mathrm{L}_B$, where $\mathrm{L}_B$ was calculated from $\mathrm{L}_R$
assuming a $B-R=0.8$, which is the average $B-R$ colour for the dwarf
galaxies in our sample (see Paper~II). For comparison, the ranges of
$M_\mathrm{\HI}/\mathrm{L}_B$ for different morphological types, as
derived by BR, are indicated by the horizontal bars.  It is clear that
the late-type dwarf galaxies have a much higher
$M_\mathrm{\HI}/\mathrm{L}_B$ than earlier type spiral galaxies, a
result already found by Roberts (1969, see also Roberts \& Haynes
1994), although for this sample the results may be biased towards
higher values because we selected galaxies to have flux densities in
excess of 200 mJy.  The average value of $M_\mathrm{\HI}/\mathrm{L}_B$
for the late-type dwarf galaxies presented here is $1.5\pm 1.0$
$\mathrm{M}_\odot/\mathrm{L}_{B,\odot}$.

In Fig.~\ref{figmhilrmu} the correlations between $\mu_0^R$, $M_R$ and
$M_\mathrm{\HI}/\mathrm{L}_R$ are shown.  In order to compare the
dwarf galaxy properties with those of more luminous spiral galaxies,
we have included the results from two other studies.  The open circles
in Fig.~\ref{figmhilrmu} are the Ursa Major galaxies from VS, and the
crosses are the low surface brightness galaxies from de Blok, McGaugh
\& van der Hulst (1996). The full circles are the data from our
sample.  A clear trend is visible between $\mu_0^R$ and
$M_\mathrm{\HI}/\mathrm{L}_R$, in the sense that lower surface
brightness galaxies are richer in \HI. Also, there appears to be a
trend between $M_R$ and $M_\mathrm{\HI}/\mathrm{L}_R$, although among
dwarf galaxies this trend is weak. Fig.~\ref{figmhilrmu}c shows the
weak correlation between $\mu_0^R$ and $M_R$.  The dwarf galaxies in
our sample have the same range in surface brightness at each absolute
magnitude. This is not true of the bright spiral galaxies in the UMa
sample of VS, which predominantly have high surface
brightnesses. Note that the data presented in Fig.~\ref{figmhilrmu} do
not represent a complete sample, and some galaxies, such as compact
high surface brightness dwarfs or extended low surface brightness
galaxies may be missing.

\begin{figure}[t]
\resizebox{\hsize}{!}{\includegraphics{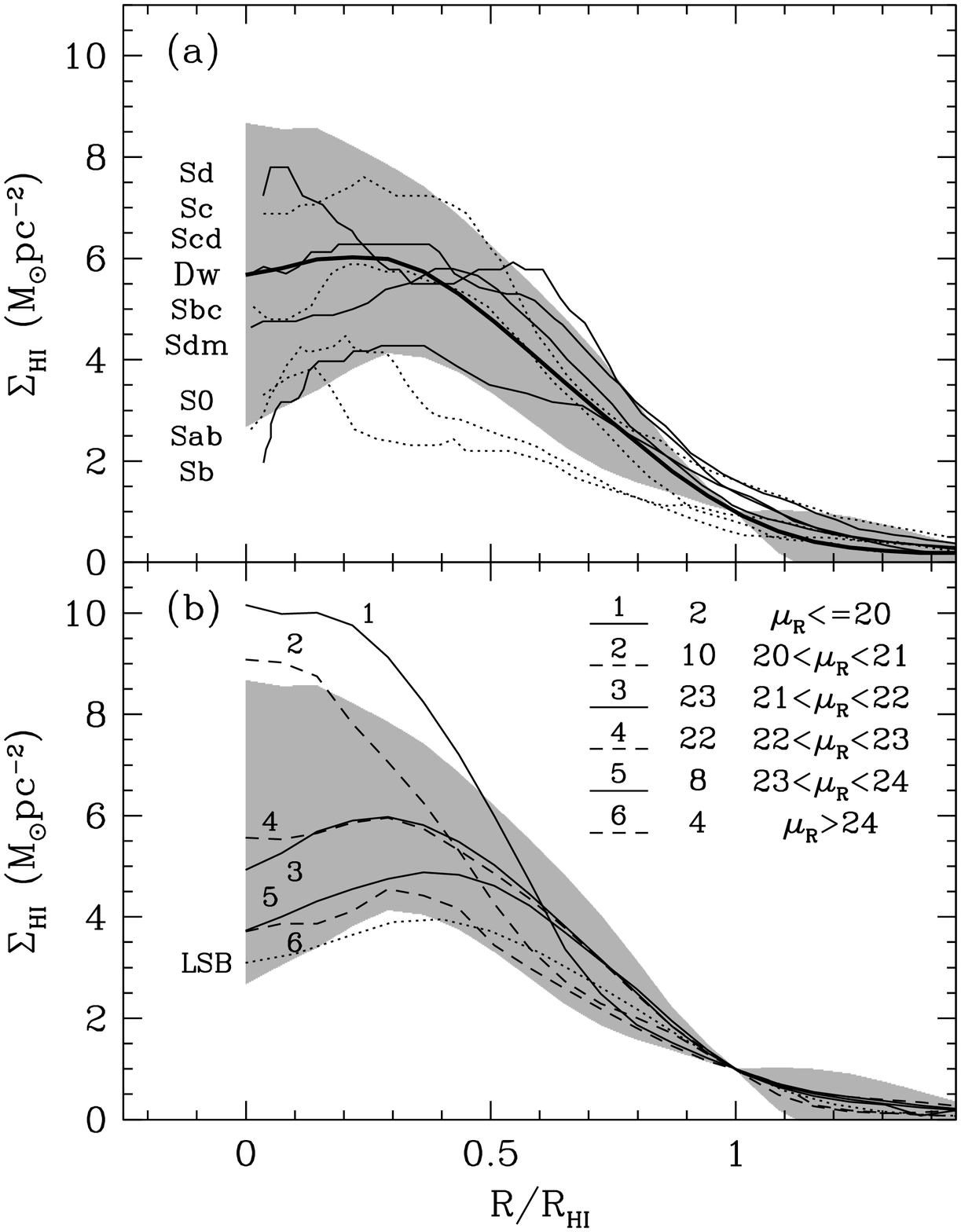}}
\vspace{-6mm}
\caption{ {\bf (a)} Comparison of the range in radial \figHI\ surface
  density profiles among dwarf galaxies and bright spiral galaxies.
  The thick black line indicated by `Dw' represents the average
  profile of all dwarf galaxies in our sample, and the shaded area
  indicates one standard deviation at each radius around the mean. The
  other lines represent average profiles for different morphological
  types from Cayatte \etal\ (1994). {\bf (b)} Radial \figHI\ surface
  density profiles of the dwarf galaxies, binned by central disk
  surface brightness as indicated in the top right. In front of the
  range in surface brightnesses the number of galaxies in each bin is
  given. The shaded area is the same as in panel (a). The dotted line
  is the average profile for LSB galaxies from de Blok \etal\ (1996).}
\label{figradialcomp}
\end{figure}

\subsection{\HI\ surface density versus surface brightness}

\begin{figure}[ht]
\resizebox{\hsize}{!}{\includegraphics{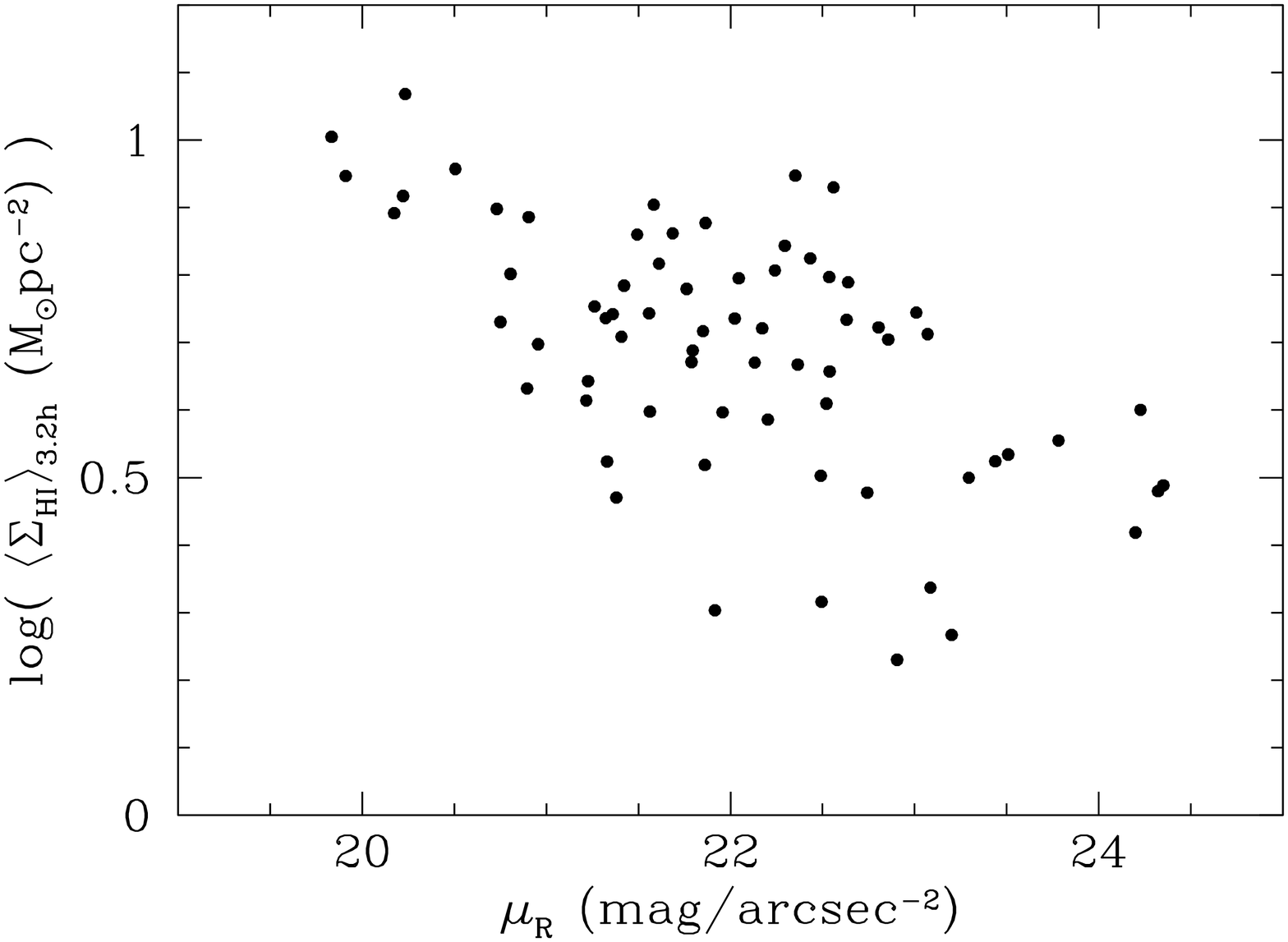}}
\caption{Average \figHI\ surface density within 3.2 disk scale lengths
  ($\langle\Sigma_\mathrm{\HI} \rangle_{3.2h}$) versus central disk
  surface brightness $\mu_0^R$. }
\label{figdens}
\end{figure}

Inspection of the radial \HI\ surface density profiles in
Fig.~\ref{figexpfits} or in Appendix~B, makes it clear that the shapes
of the radial \HI\ surface density profiles range from centrally
peaked profiles to profiles with central holes. In
Fig.~\ref{figradialcomp}a the radial density profiles for the galaxies
in our sample are compared to the average radial profiles compiled by
Cayatte \etal\ (1994, hereafter CKBG).  These authors constructed a
comparison sample of radial profiles of undisturbed galaxies based on
Warmels (1988c) and Broeils (1992a), as a comparison sample for their
Virgo cluster galaxies.  CKBG give their profiles in units of
$R_{25}$.  We converted their profiles to units of $R_\mathrm{\HI}$,
using the mean $R_\mathrm{\HI}/R_{25}$ given in CKBG.  Because it is
confusing to plot all our 73 radial profiles in one figure, the
average of all the profiles is shown with the thick line. The shaded
area represents one standard deviation at each radius around the mean
surface brightness. We note that our selection on flux densities
larger than 200 mJy may have introduced a bias towards systems that
are rich in \HI.

The comparison with the profiles compiled by CKBG shows that the
radial profiles found in dwarf galaxies indeed cover a large range
shapes, from profiles seen in early-type spirals to
those seen in late-type spiral galaxies. Because there is an overlap
in morphological types between our sample and late-type spirals, it is
not surprising that some of the radial profiles are similar.  On the
other hand, it is surprising to see that a significant fraction of
dwarf galaxies have low \HI\ densities and radial profiles like those
found in early-type spiral galaxies. The early type spiral galaxies
are believed to have consumed most of their \HI\ and hence have a low
\HI\ content. The low \HI\ density dwarf galaxies, on the other hand,
are rich in \HI, as measured by $M_\mathrm{\HI}/\mathrm{L}_R$ (see
Fig.~\ref{figmhilrmu}).

A clue as to why a significant fraction of the dwarf galaxies have low
\HI\ densities is provided in Fig.~\ref{figradialcomp}b. In this
figure, the radial \HI\ density profiles are shown averaged in surface
brightness bins. It is clear that galaxies with high optical surface
brightnesses have higher \HI\ densities as well. The dwarfs with the
lowest surface brightnesses, $\mu_R>23$, have radial \HI\ profiles
much like those of the LSB galaxies studied by de Blok \etal\
(1996), shown by the dotted line. These LSB galaxies have
central surface brightnesses in the range $22<\mu_R<24$. It seems that
the \HI\ densities and the luminosity densities are coupled, as was
also found by de Blok \etal\ (1996).

This coupling is explored further in Fig.~\ref{figdens}, where the
correlation between the average \HI\ density within 3.2 disk scale
lengths ($\langle\Sigma_\mathrm{\HI} \rangle_{3.2h}$) and the central
surface brightness is shown. The central surface brightness is
directly related to the average surface brightness for an exponential
disk. Because most of the dwarf galaxies have light profiles close to
purely exponential (see Paper~II), the central surface brightness has
been used in Fig.~\ref{figdens}.  It is clear that galaxies with
higher surface brightnesses have higher \HI\ densities as well.
However, the \HI\ density changes much more slowly than the surface
brightness in the $R$-band. Over a range of 4 magnitudes in surface
brightness, i.e., a factor of 40 in luminosity density, the \HI\ 
density changes only by about a factor of 4.  Therefore, it seems that
these galaxies have low \HI\ densities not because they have consumed
their \HI, like the early type spiral galaxies may have, but because their
baryonic mass density in general is lower.

We also investigated whether a correlation exists between \HI\ radial
profile shape and absolute magnitude. No such correlation was found;
at each absolute magnitude, all profile shapes occur.

\section{\HI\ morphology}
\label{themorphology}

Large-scale asymmetries in the optical appearance of galaxies have
been known for a long time. But galaxies can also be asymmetric in
their \HI\ distributions. Baldwin \etal\ (1980) drew
attention to lopsided \HI\ distributions of disk galaxies, emphasizing
that the asymmetry affects large parts of the disk and that it is a
common phenomenon among spiral galaxies.  Richter \& Sancisi (1994)
made an estimate of the frequency of asymmetries from the shape of the
\HI\ line profiles. From an inspection of about 1700 global profiles
they found that at least half of the disk galaxies have strong or mild
asymmetries. This result was confirmed by Haynes \etal\ (1998).
Swaters \etal\ (1999) found that lopsidedness may
not only be seen in the morphology, but also in the kinematics. They
find that probably at least half of all galaxies are kinematically
lopsided.

With the present sample of 73 late-type dwarf galaxies, we can
investigate the frequency of lopsidedness among dwarf galaxies. To
this end, we have inspected the data presented in Appendix~B
by eye, and looked for lopsidedness in the global profiles, the
integrated \HI\ maps and the velocity fields. The results are listed
in Table~\ref{tabprops}.

We find that lopsidedness is as common among dwarf galaxies as it is
among spiral galaxies. We find that 25 of the 73 global profiles are
clearly asymmetric, and another 12 show mild asymmetries, bringing the
total fraction of asymmetric global profiles to about 50\%, similar to
the fraction seen among spiral galaxies (Richter \& Sancisi 1994). The
fraction of dwarf galaxies that are lopsided in their \HI\
distribution is about 35\%. The fraction of kinematically lopsided
dwarf galaxies has been determined from the appearance of the
position-velocity diagrams and the velocity fields presented in
Appendix~B. For 19 galaxies we found that the signal-to-noise ratio
was too low to determine whether a lopsidedness in the kinematics was
present. For the remaining 54 galaxies, we found that 16 show clear
signs of kinematic lopsidedness, and 11 show weak signs, bringing the
total at about 50\%, concordant with the fraction found in Swaters
\etal\ (1999).

In five of the galaxies the lopsidedness is clearly due to ongoing
interaction (\object{UGC~1249}, \object{UGC~5272}, \object{UGC~5935},
\object{UGC~6944} and \object{UGC~7592}). In the other cases, there is
no clear present interaction. Many of the dwarf galaxies in our sample
are members of small groups, and hence are likely to have undergone
interaction in the past. The distances to the nearest companions may
be large, however, and therefore the excited lopsidedness must be
long-lived to explain the high observed fractions of asymmetries in
the global profiles, the \HI\ distribution and the kinematics.

\subsection{Notes on individual galaxies}
\label{thenotes}

\notes{\object{UGC~1249}} is interacting with the nearby spiral galaxy
\object{UGC~1256} which explains its distorted kinematics. The \HI\ to
the NE of \object{UGC~1249} is part of a bridge between the two
galaxies. We note that these data are somewhat affected by
instrumental effects.

\notes{\object{UGC~4274}} is also known as the Bearpaw Galaxy. The
\HI\ in this galaxy is concentrated in one clump that is offset from
the optical center.  This galaxy was observed in the same pointing as
UGC~4278.  Because \object{UGC~4274} was far away from the pointing
center, the noise at its position was higher after correction for the
primary beam attenuation.  This explains the large difference between
the gray scale and the three sigma contours in the figure in
Appendix~B.

\notes{\object{UGC~4305}} shows many holes in its \HI\ distribution.
The morphology and kinematics of this galaxy, \object{Holmberg~II},
have been studies in detail by Puche \etal\ (1992).

\notes{\object{UGC~5272}} has a small companion $2'$ to the south,
with an \HI\ mass of about $1\cdot 10^7$ \msun\ and an absolute
magnitude of $M_R\approx -12.5$. \object{UGC~5272} and its companion
are connected by an \HI\ bridge.

\notes{\object{UGC~5721}} is strongly lopsided in its \HI\ 
distribution, even though it appears isolated on the sky.

\notes{\object{UGC~5935}} is interacting with \object{UGC~5931}. The
extended features in the figure in Appendix~B do not represent tidal
\HI\, but are the result of instrumental effects.

\notes{\object{UGC~5986}} has a strong warp on the SW end. This warp
may be related to a small companion, visible in the optical image at
the position where the warp sets in.  The \HI\ on the NE end appears
disrupted as well, with an extension in the same direction as the warp
on the SW end. The possible companion does not have \HI\ emission and
its redshift is not catalogued, and thus it may be a background
galaxy.

\notes{\object{UGC~6944}} is part of a small group consisting of three
bright galaxies. Two of the members contain \HI\ (\object{UGC~6933}
and \object{UGC~6944}), one has no \HI, even though its optically
determined systemic velocity falls within the velocity range of these
observations. There is some \HI\ seen between the galaxies, most
likely as the result of interactions between them.

\notes{\object{UGC~7592}} has a huge, low column density \HI\ 
envelope. The \HI\ in the optical part of the galaxy appears to rotate
in the opposite sense as the outer \HI, although it may also be that
the outer \HI\ is warped through the plane of the sky. The properties
of the outer \HI\ are studied in more detail in Hunter \etal\ (1998).

\notes{\object{UGC~8490}} has a strong warp that is already visible
from the morphology of the system, but the kinematics show it more
clearly.

\section{Conclusions}
\label{thesummary}

From the neutral hydrogen observations for the sample of 73 late-type
dwarf galaxies presented here, we obtain the following results.

\noindent (1) The ratio of the \HI\ extent to the optical diameter,
defined as 6.4 disk scale lengths, is on average $1.8\pm 0.8$, similar
to the value found for spiral galaxies, but with a larger spread.

\noindent (2) Most of the dwarf galaxies in this sample are rich in
\HI\, with typical $M_\mathrm{\HI}/L_B$ values of 1.5.

\noindent (3) The relative \HI\ content $M_\mathrm{\HI}/L_B$
increases towards fainter absolute magnitudes and towards fainter
surface brightnesses.

\noindent (4) Dwarf galaxies with lower average surface brightnesses
also have lower average \HI\ column densities. Over a range of 4
magnitudes in surface brightness, i.e., a factor of 40 in luminosity
density, the \HI\ density changes only by about a factor of 4.

\noindent (5) We find that lopsidedness is as common among dwarf
galaxies as it is in spiral galaxies. About half of the dwarf galaxies
in our sample have asymmetric global profiles, a third has a lopsided
\HI\ distribution, and about half shows signs of kinematic
lopsidedness.

\acknowledgements

Dolf Sijbring and Jurjen Kamphuis are acknowledged for their work on
the reduction of the WHISP data.  We thank Liese van Zee for providing
optical $R$-band images for \object{UGC~10310} and \object{UGC~11861}.
The WSRT is operated by the Netherlands Foundation for Research in
Astronomy with financial support from the Netherlands Organization for
Scientific Research (NWO). This research has made use of the NASA/IPAC
Extragalactic Database (NED) which is operated by the Jet Propulsion
Laboratory, California Institute of Technology, under contract with
the National Aeronautics and Space Administration.

\endacknowledgements

\clearpage

\appendix

\section{Tables}
\label{thetables}

\subsection[The sample]{Table~\ref{tabsample} -- The sample}

\noindent {\it Column} (1) gives the UGC number. For a description of
the sample selection, see Sect.~\ref{thesample}.

\noindent {\it Column} (2) provides other common names, in this order:
NGC, DDO (van den Bergh 1959, 1966), IC, Arp (Arp 1966), CGCG.  At most
two other names are given.

\noindent {\it Columns} (3) and (4) give the equatorial coordinates
(1950) derived from the optical images, as described in
Paper~II.

\noindent {\it Column} (5) gives the morphological type according to
the RC3, using the same coding.

\noindent {\it Column} (6) provides the adopted distance.  Where   
possible, stellar distance indicators have been used, mostly Cepheids
and brightest stars.  If these were not available, a distance based on
group membership was used.  If this was not available either, the
distance was calculated from the \HI\ systemic velocity following the
prescription given in Kraan-Korteweg (1986), with an adopted Hubble
constant of $H_0=75$ \kms\ Mpc$^{-1}$.  A full list of published
distances for the galaxies in this sample, updated to the beginning of
1998, is given in Table~A2 of Paper~II. A discussion on the
distance uncertainties is given in Sect.~3 of Paper~II.

\noindent {\it Column} (7) gives the absolute $B$-band magnitude,
calculated from the apparent photographic magnitude as given in the
RC3, and the distance as given in column 6.

\noindent {\it Column} (8) lists the extrapolated central $R$-band
disk surface brightnesses as determined from fits to the surface
brightness profiles presented in Paper~II. The values have
been corrected for Galactic foreground extinction (derived from the
$A_B$ value according to Burstein \& Heiles (1984) assuming $A_B/
A_R$ of 1.77 (Rieke \& Lebofsky 1985)), and were corrected to face-on,
assuming that the galaxies are transparent.

\noindent {\it Column} (9) gives the $R$-band disk scale length, as
determined from the surface brightness profiles presented in
Paper~II.

\noindent {\it Column} (10) gives the diameter at which the 25
$R$-band mag arcsec$^{-2}$ is reached, after correction for Galactic
foreground extinction and inclination.

\subsection[\HI\ properties]{Table~\ref{tabprops} -- \HI\ properties}

\noindent {\it Column} (1) gives the UGC number.

\noindent {\it Column} (2) lists the systemic heliocentric velocity.

\noindent {\it Column} (3) and (4) give the linewidths as determined
from the global profile, corrected for random motions and
inclinations. Column 3 gives the linewidth at the 20\% level, column 4
at the 50\% level.

\noindent {\it Column} (5) contains the integrated \HI\ fluxes derived
from the global profiles.

\noindent {\it Column} (6) lists the \HI\ mass in units of $10^8$
\msun.

\noindent {\it Column} (7) gives the \HI\ radius, defined as the
radius where the \HI\ surface density corrected to face-on reaches 1
\msun pc$^{-2}$.

\noindent {\it Column} (8) gives the \HI\ scale length, as determined
from a fit to the outer parts of the radial \HI\ density profile.

\noindent {\it Column} (9) lists the average \HI\ surface density within
3.2 disk scale lengths $\langle\Sigma_\mathrm{\HI}\rangle_{3.2h}$.

\noindent {\it Column} (10), (11) and (12) indicate whether a galaxy
was found to be lopsided in the global profile, the \HI\ distribution
or the kinematics. An single star (\star) indicates weak lopsidedness,
and a double star (\star\star) strong lopsidedness.

\clearpage

\setlength{\tabcolsep}{4.5pt}
\def\ruleit#1{\vrule width#1 height0pt depth0pt}
{
\begin{table*}
{\vspace{-0.2cm}}
\caption[]{The sample}\label{tabsample}\null
\begin{flushleft}
\begin{tabular}{rlrrrrrrcrrrrr}
\noalign{\vspace{-0.2cm}}\hline\noalign{\smallskip}
 UGC & \hfil Other names & \multicolumn{3}{c}{R.A. (1950)}& \multicolumn{3}{c}{Dec. (1950)} & Type & D$_a$ & $M_B$ & $\mu_0^R$ & $h$ & $D_{25}^{B,i}$ \\
  &  & \multicolumn{3}{c}{$^h$ $^m$ $^s$}& \multicolumn{3}{c}{$^\circ$ $'$ $''$} &  & Mpc & mag & \hbox to20pt{mag/${\prime\prime}^2$} & $''$ & $''$ \\
 (1) & \hfil (2) & \multicolumn{3}{c}{(3)}& \multicolumn{3}{c}{(4)} & (5) & (6) & (7) & (8) & (9) & (10) \\
\noalign{\smallskip}\hline\noalign{\smallskip}
\omit\ruleit{1.0cm}&\ruleit{3.4cm}&\ruleit{0.4cm}&\ruleit{0.4cm}&\ruleit{0.6cm}&\ruleit{0.6cm}&\ruleit{0.4cm}&\ruleit{0.4cm}&\ruleit{1.5cm}&\ruleit{0.7cm}&\ruleit{1.0cm}&\ruleit{0.9cm}&\ruleit{0.9cm}&\ruleit{0.9cm}\\
731 & DDO 9 & 1 & 07 & 46.7 & 49 & 20 & 7 & .I..9*. & 8.0 & -16.6 & 23.0 & 46 & 92 \\
1249 & IC 1727 & 1 & 44 & 40.9 & 27 & 04 & 59 & .SBS9.. & 7.5 & -17.9 & 22.1 & 56 & 143 \\
1281 &  & 1 & 46 & 38.9 & 32 & 20 & 31 & .S..8.. & 5.5 & -16.2 & 22.7 & 46 & 85 \\
2023 & DDO 25 & 2 & 30 & 17.4 & 33 & 16 & 18 & .I..9*. & 10.1 & -17.2 & 21.8 & 25 & 76 \\
2034 & DDO 24 & 2 & 30 & 34.4 & 40 & 18 & 34 & .I..9.. & 10.1 & -17.5 & 21.6 & 26 & 78 \\
2053 & DDO 26 & 2 & 31 & 32.0 & 29 & 31 & 57 & .I..9.. & 11.8 & -16.0 & 22.5 & 19 & 43 \\
2455 & NGC 1156 & 2 & 56 & 46.5 & 25 & 02 & 21 & .IBS9.. & 7.8 & -18.5 & 19.8 & 23 & 111 \\
3137 &  & 4 & 39 & 21.3 & 76 & 19 & 35 & .S?.... & 18.4 & -18.7 & 24.2 & 65 & 48 \\
3371 & DDO 39 & 5 & 49 & 49.1 & 75 & 18 & 30 & .I..9*. & 12.8 & -17.7 & 23.3 & 53 & 81 \\
3698 &  & 7 & 05 & 42.5 & 44 & 27 & 40 & .I..9*. & 8.5 & -15.4 & 21.2 & 10 & 37 \\
3711 & NGC 2337 & 7 & 06 & 37.2 & 44 & 32 & 21 & .IB.9.. & 8.6 & -17.8 & 20.9 & 22 & 81 \\
3817 &  & 7 & 19 & 07.9 & 45 & 12 & 18 & .I..9*. & 8.7 & -15.1 & 22.5 & 16 & 36 \\
3851 & NGC 2366, DDO 42 & 7 & 23 & 35.1 & 69 & 18 & 53 & .IBS9.. & 3.4 & -16.9 & 22.6 & 88 & 207 \\
3966 & DDO 46 & 7 & 38 & 01.3 & 40 & 13 & 41 & .I..9.. & 6.0 & -14.9 & 22.2 & 19 & 47 \\
4173 &  & 7 & 59 & 04.6 & 80 & 16 & 10 & .I..9*. & 16.8 & -17.8 & 24.3 & 61 & 35 \\
4274 & NGC 2537, Arp 6 & 8 & 09 & 42.8 & 46 & 08 & 28 & .SBS9P. & 6.6 & -18.0 & 20.7 & 23 & 91 \\
4278 & IC 2233 & 8 & 10 & 27.5 & 45 & 53 & 43 & .SBS7*/ & 10.5 & -17.7 & 22.5 & 45 & 78 \\
4305 & Arp 268 & 8 & 13 & 54.9 & 70 & 52 & 46 & .I..9.. & 3.4 & -16.8 & 21.7 & 60 & 178 \\
4325 & NGC 2552 & 8 & 15 & 40.1 & 50 & 09 & 57 & .SAS9\$. & 10.1 & -18.1 & 21.6 & 36 & 105 \\
4499 &  & 8 & 34 & 01.9 & 51 & 49 & 39 & .SX.8.. & 13.0 & -17.8 & 21.5 & 22 & 70 \\
4543 &  & 8 & 39 & 55.7 & 45 & 54 & 59 & .SA.8.. & 30.3 & -19.2 & 22.0 & 23 & 61 \\
5272 & DDO 64 & 9 & 47 & 26.8 & 31 & 43 & 16 & .I..9.. & 6.1 & -15.1 & 22.4 & 21 & 49 \\
5414 & NGC 3104, Arp 264 & 10 & 00 & 56.3 & 40 & 59 & 57 & .IXS9.. & 10.0 & -17.6 & 21.8 & 30 & 89 \\
5721 & NGC 3274 & 10 & 29 & 30.3 & 27 & 55 & 35 & .SX.7?. & 6.7 & -16.6 & 20.2 & 14 & 62 \\
5829 & DDO 84 & 10 & 39 & 54.0 & 34 & 42 & 45 & .I..9.. & 9.0 & -17.3 & 22.4 & 39 & 94 \\
5846 & DDO 86 & 10 & 41 & 17.2 & 60 & 37 & 52 & .I..9.. & 13.2 & -16.1 & 22.9 & 19 & 36 \\
5918 & VII Zw 347 & 10 & 46 & 17.6 & 65 & 47 & 41 & .I..9*. & 7.7 & -15.4 & 24.2 & 46 & 27 \\
5935 & NGC 3396, Arp 270 & 10 & 47 & 08.3 & 33 & 15 & 20 & .IB.9P. & 26.4 & -20.1 & 21.8 & 31 & 93 \\
5986 & NGC 3432, Arp 206 & 10 & 49 & 42.8 & 36 & 53 & 6 & .SBS9./ & 8.7 & -18.6 & 21.4 & 46 & 149 \\
6446 &  & 11 & 23 & 52.9 & 54 & 01 & 20 & .SA.7.. & 12.0 & -18.3 & 21.4 & 28 & 95 \\
6628 &  & 11 & 37 & 25.7 & 46 & 13 & 10 & .SA.9.. & 15.3 & -18.8 & 21.8 & 36 & 100 \\
6817 & DDO 99 & 11 & 48 & 18.0 & 39 & 09 & 34 & .I..9.. & 4.02 & -15.2 & 23.1 & 48 & 84 \\
6944 & NGC 3995, Arp 313 & 11 & 55 & 09.8 & 32 & 34 & 19 & .SA.9P. & 47.4 & -21.2 & 20.4 & 17 & 76 \\
6956 & DDO 102 & 11 & 55 & 51.4 & 51 & 11 & 48 & .SBS9.. & 15.7 & -17.2 & 23.4 & 33 & 51 \\
7047 & NGC 4068 & 12 & 01 & 29.6 & 52 & 52 & 7 & .IA.9.. & 3.5 & -15.2 & 21.6 & 27 & 81 \\
7125 &  & 12 & 06 & 10.0 & 37 & 04 & 51 & .S..9.. & 19.5 & -18.3 & 22.8 & 34 & 76 \\
7151 & NGC 4144 & 12 & 07 & 27.2 & 46 & 44 & 9 & .SXS6\$/ & 3.5 & -15.7 & 22.3 & 44 & 121 \\
7199 & NGC 4163 & 12 & 09 & 37.6 & 36 & 26 & 47 & .IA.9.. & 3.5 & -15.1 & 21.4 & 22 & 73 \\
7232 & NGC 4190 & 12 & 11 & 13.6 & 36 & 54 & 49 & .I..9P. & 3.5 & -15.3 & 20.2 & 15 & 68 \\
7261 & NGC 4204 & 12 & 12 & 42.1 & 20 & 56 & 14 & .SBS8.. & 9.1 & -17.7 & 21.9 & 35 & 102 \\
7278 & NGC 4214 & 12 & 13 & 08.6 & 36 & 36 & 17 & .IXS9.. & 3.5 & -18.3 & 20.2 & 53 & 237 \\
7323 & NGC 4242 & 12 & 15 & 01.3 & 45 & 53 & 49 & .SXS8.. & 8.1 & -18.9 & 21.2 & 54 & 176 \\
7399 & NGC 4288, DDO 119 & 12 & 18 & 10.4 & 46 & 34 & 9 & .SBS8.. & 8.4 & -17.1 & 20.7 & 18 & 70 \\
7408 & DDO 120 & 12 & 18 & 47.7 & 46 & 05 & 25 & .IA.9.. & 8.4 & -16.6 & 21.9 & 24 & 70 \\
7490 & DDO 122 & 12 & 22 & 10.4 & 70 & 36 & 39 & .SA.9.. & 8.5 & -17.4 & 21.3 & 27 & 90 \\
7524 & NGC 4395 & 12 & 23 & 19.9 & 33 & 49 & 26 & .SAS9*. & 3.5 & -18.1 & 22.2 & 135 & 372 \\
7559 & DDO 126 & 12 & 24 & 37.5 & 37 & 25 & 9 & .IB.9.. & 3.2 & -13.7 & 23.8 & 45 & 49 \\
7577 & DDO 125 & 12 & 25 & 15.7 & 43 & 46 & 13 & .I..9.. & 3.5 & -15.6 & 22.5 & 51 & 115 \\
7592 & NGC 4449 & 12 & 25 & 45.2 & 44 & 22 & 11 & .IB.9.. & 3.5 & -18.5 & 20.3 & 51 & 215 \\
7603 & NGC 4455 & 12 & 26 & 13.8 & 23 & 05 & 53 & .SBS7?/ & 6.8 & -16.9 & 20.8 & 21 & 79 \\
7608 & DDO 129 & 12 & 26 & 18.7 & 43 & 30 & 7 & .I..9.. & 8.4 & -16.4 & 22.6 & 30 & 68 \\
7690 &  & 12 & 30 & 01.6 & 42 & 58 & 49 & .I..9*. & 7.9 & -17.0 & 19.9 & 12 & 61 \\
\noalign{\smallskip}\hline
\end{tabular}
\end{flushleft}
\end{table*}
\addtocounter{table}{-1}
\begin{table*}
{\vspace{-0.2cm}}
\caption[]{-- Continued\\}
\vskip-\baselineskip
\begin{flushleft}
\begin{tabular}{rlrrrrrrcrrrrr}
\noalign{\vspace{-0.2cm}}\hline\noalign{\smallskip}
 UGC & \hfil Other names & \multicolumn{3}{c}{R.A. (1950)}& \multicolumn{3}{c}{Dec. (1950)} & Type & D$_a$ & $M_B$ & $\mu_0^R$ & $h$ & $D_{25}^{B,i}$ \\
  &  & \multicolumn{3}{c}{$^h$ $^m$ $^s$}& \multicolumn{3}{c}{$^\circ$ $'$ $''$} &  & Mpc & mag & \hbox to20pt{mag/${\prime\prime}^2$} & $''$ & $''$ \\
 (1) & \hfil (2) & \multicolumn{3}{c}{(3)}& \multicolumn{3}{c}{(4)} & (5) & (6) & (7) & (8) & (9) & (10) \\
\noalign{\smallskip}\hline\noalign{\smallskip}
\omit\ruleit{1.0cm}&\ruleit{3.4cm}&\ruleit{0.4cm}&\ruleit{0.4cm}&\ruleit{0.6cm}&\ruleit{0.6cm}&\ruleit{0.4cm}&\ruleit{0.4cm}&\ruleit{1.5cm}&\ruleit{0.7cm}&\ruleit{1.0cm}&\ruleit{0.9cm}&\ruleit{0.9cm}&\ruleit{0.9cm}\\
7866 & IC 3687 & 12 & 39 & 50.8 & 38 & 46 & 39 & .IXS9.. & 4.8 & -15.2 & 22.1 & 25 & 69 \\
7916 & I Zw 42 & 12 & 41 & 59.9 & 34 & 39 & 37 & .I..9.. & 8.4 & -14.9 & 24.4 & 42 & 26 \\
7971 & NGC 4707, DDO 150 & 12 & 46 & 05.9 & 51 & 26 & 16 & .S..9*. & 8.4 & -17.1 & 21.3 & 23 & 77 \\
8188 & IC 4182 & 13 & 03 & 30.2 & 37 & 52 & 27 & .SAS9.. & 4.7 & -17.4 & 21.3 & 52 & 170 \\
8201 & VII Zw 499 & 13 & 04 & 38.2 & 67 & 58 & 21 & .I..9.. & 4.9 & -15.8 & 21.9 & 32 & 96 \\
8286 & NGC 5023 & 13 & 09 & 58.0 & 44 & 18 & 11 & .S..6*/ & 4.8 & -17.2 & 20.9 & 34 & 124 \\
8331 & DDO 169 & 13 & 13 & 19.8 & 47 & 45 & 49 & .IA.9.. & 5.9 & -15.1 & 22.9 & 28 & 54 \\
8490 & NGC 5204 & 13 & 27 & 43.9 & 58 & 40 & 39 & .SAS9.. & 4.9 & -17.3 & 20.5 & 29 & 128 \\
8550 & NGC 5229 & 13 & 31 & 58.6 & 48 & 10 & 14 & .SBS7?/ & 5.3 & -15.6 & 22.0 & 24 & 72 \\
8683 & DDO 182 & 13 & 40 & 23.2 & 39 & 54 & 33 & .I..9.. & 12.6 & -16.7 & 22.5 & 21 & 42 \\
8837 & DDO 185 & 13 & 52 & 56.0 & 54 & 08 & 51 & .IBS9./ & 5.1 & -15.7 & 23.2 & 50 & 79 \\
9128 & DDO 187 & 14 & 13 & 38.9 & 23 & 17 & 12 & .I..9.. & 4.4 & -14.3 & 21.9 & 17 & 50 \\
9211 & DDO 189 & 14 & 20 & 38.0 & 45 & 36 & 37 & .I..9*. & 12.6 & -16.2 & 22.6 & 19 & 42 \\
9992 &  & 15 & 41 & 26.0 & 67 & 24 & 44 & .I..9.. & 10.4 & -15.9 & 22.2 & 16 & 39 \\
10310 & Arp 2 & 16 & 14 & 49.1 & 47 & 10 & 7 & .SBS9.. & 15.6 & -17.9 & 22.0 & 25 & 70 \\
11557 &  & 20 & 23 & 01.2 & 60 & 01 & 54 & .SXS8.. & 23.8 & -19.7 & 21.0 & 26 & 80 \\
11707 &  & 21 & 12 & 20.3 & 26 & 31 & 36 & .SA.8.. & 15.9 & -18.6 & 23.1 & 58 & 94 \\
11861 &  & 21 & 55 & 44.0 & 73 & 01 & 20 & .SX.8.. & 25.1 & -20.8 & 21.4 & 53 & 148 \\
12060 &  & 22 & 28 & 17.4 & 33 & 33 & 50 & .IB.9.. & 15.7 & -17.9 & 21.6 & 21 & 73 \\
12632 & DDO 217 & 23 & 27 & 33.0 & 40 & 42 & 55 & .S..9*. & 6.9 & -17.1 & 23.5 & 85 & 120 \\
12732 &  & 23 & 38 & 09.0 & 25 & 57 & 33 & .S..9*. & 13.2 & -18.0 & 22.4 & 35 & 83 \\
\noalign{\smallskip}\hline
 ~ & & & & & & & & & & \\
 ~ & & & & & & & & & & \\
 ~ & & & & & & & & & & \\
 ~ & & & & & & & & & & \\
 ~ & & & & & & & & & & \\
 ~ & & & & & & & & & & \\
 ~ & & & & & & & & & & \\
 ~ & & & & & & & & & & \\
 ~ & & & & & & & & & & \\
 ~ & & & & & & & & & & \\
 ~ & & & & & & & & & & \\
 ~ & & & & & & & & & & \\
 ~ & & & & & & & & & & \\
 ~ & & & & & & & & & & \\
 ~ & & & & & & & & & & \\
 ~ & & & & & & & & & & \\
 ~ & & & & & & & & & & \\
 ~ & & & & & & & & & & \\
 ~ & & & & & & & & & & \\
 ~ & & & & & & & & & & \\
 ~ & & & & & & & & & & \\
 ~ & & & & & & & & & & \\
 ~ & & & & & & & & & & \\
 ~ & & & & & & & & & & \\
 ~ & & & & & & & & & & \\
 ~ & & & & & & & & & & \\
 ~ & & & & & & & & & & \\
 ~ & & & & & & & & & & \\
 ~ & & & & & & & & & & \\
 ~ & & & & & & & & & & \\
 ~ & & & & & & & & & & \\
 ~ & & & & & & & & & & \\
\end{tabular}
\end{flushleft}
\end{table*}
}

\newpage\clearpage

\setlength{\tabcolsep}{3pt}
\def\ruleit#1{\vrule width#1 height0pt depth0pt}
{
\begin{table*}
{\vspace{-0.2cm}}
\caption[]{HI properties\\}\label{tabprops}
\begin{flushleft}
\begin{tabular}{rcccccccccccc}
\noalign{\vspace{-0.2cm}}\hline\noalign{\smallskip}
 UGC & $\varv_\mathrm{sys}$ & $W_{R,20}^i$ & $W_{R,50}^i$ & $\int$Sdv & $M_\mathrm{HI}$ & $R_\mathrm{HI}$ & $h_\mathrm{HI}$ & $\langle\Sigma_\mathrm{HI}\rangle_{3.2h}$ & & \multicolumn{3}{c}{lopsidedness} \\
 & \kms & \kms & \kms & Jy \kms & $10^8$ \msun & $''$ & $''$ & M$_\odot$pc$^{-2}$ & & prof. & dens. & kin.\\
 (1) & (2) & (3) & (4) & (5) & (6) & (7) & (8) & (9) & & (10) & (11) & (12) \\
\noalign{\smallskip}\hline\noalign{\smallskip}
\omit\ruleit{1.2cm}&\ruleit{0.9cm}&\ruleit{0.9cm}&\ruleit{0.9cm}&\ruleit{1.05cm}&\ruleit{0.9cm}&\ruleit{0.9cm}&\ruleit{0.9cm}&\ruleit{0.9cm}&\ruleit{0.4cm}&\ruleit{0.9cm}&\ruleit{0.9cm}&\ruleit{0.9cm}\\
731 & 639 & 143 & 145 & 48.8 &  7.4 & 191 & 36 &  5.5 & & \star\star &  & \star\star \\
1249 & 340 & 129 & 144 & 23.5 &  3.1 & 362 & 135 &  4.3 & & \star\star & \star\star & \star\star \\
1281 & 157 & 112 & 113 & 44.4 &  3.2 & 206 & 60 &  3.0 & &  &  &  \\
2023 & 603 & 85 & 110 & 18.7 &  4.5 & 129 & 31 &  4.7 & &  &  &  \\
2034 & 578 & 103 & 134 & 35.8 &  8.6 & 189 & 29 &  5.5 & &  &  &  \\
2053 & 1029 & 79 & 101 & 17.1 &  5.6 & 120 & 28 &  6.3 & & \star &  &  \\
2455 & 380 & 80 & 118 & 71.3 & 10.2 & 212 & 54 & 10.1 & &  &  &  \\
3137 & 993 & 218 & 216 & 54.5 & 43.6 & 297 & 95 &  4.0 & & \star\star &  &  \\
3371 & 816 & 155 & 159 & 31.5 & 12.2 & 188 & 31 &  3.2 & & \star\star &  & \star \\
3698 & 426 & 57 & 63 &  6.7 &  1.1 & 64 & 39 &  4.4 & & \star\star &  &  \\
3711 & 433 & 160 & 177 & 39.4 &  6.9 & 164 & 62 &  7.7 & &  &  &  \\
3817 & 438 & 72 & 87 & 12.9 &  2.3 & 103 & 33 &  4.5 & &  &  &  \\
3851 & 99 & 106 & 111 & 267.4 &  7.3 & 439 & 118 &  8.5 & & \star\star & \star\star & \star\star \\
3966 & 361 & 112 & 117 & 24.8 &  2.1 & 135 & 56 &  6.4 & & \star &  &  \\
4173 & 862 & 95 & 108 & 31.8 & 21.2 & 178 & 34 &  3.0 & & \star & \star\star &  \\
4274 & 447 & 132 & 132 & 14.8 &  1.5 & 126 & 40 &  5.4 & & \star\star & \star\star &  \\
4278 & 561 & 161 & 167 & 52.3 & 13.6 & 193 & 32 &  4.1 & & \star &  & \star \\
4305 & 158 & 79 & 89 & 246.8 &  6.7 & 443 & 93 &  7.3 & &  &  &  \\
4325 & 519 & 185 & 189 & 31.2 &  7.5 & 142 & 25 &  6.6 & & \star\star & \star & \star\star \\
4499 & 691 & 138 & 142 & 29.8 & 11.9 & 143 & 31 &  7.2 & &  &  & \star \\
4543 & 1960 & 140 & 157 & 34.0 & 73.6 & 144 & 71 &  5.4 & &  & \star &  \\
5272 & 520 & 79 & 102 & 19.3 &  1.7 & 106 & 22 &  8.9 & &  & \star\star &  \\
5414 & 612 & 117 & 123 & 27.4 &  6.5 & 146 & 35 &  6.0 & & \star\star & \star & \star\star \\
5721 & 537 & 167 & 169 & 62.6 &  6.6 & 225 & 65 & 11.7 & & \star\star & \star\star &  \\
5829 & 629 & 119 & 142 & 58.2 & 11.1 & 188 & 35 &  6.7 & & \star &  &  \\
5846 & 1019 & 83 & 95 & 17.4 &  7.1 & 133 & 31 &  5.1 & &  &  &  \\
5918 & 338 & 78 & 88 & 21.1 &  3.0 & 159 & 43 &  2.6 & &  & \star\star &  \\
5935 & 1633 & 163 & 178 & 62.8 & 103.3 & 144 & 104 &  5.0 & & \star\star & \star\star & \star\star \\
5986 & 616 & 229 & 238 & 151.8 & 27.1 & 395 & 60 &  6.1 & & \star\star & \star\star & \star\star \\
6446 & 645 & 152 & 160 & 39.7 & 13.5 & 182 & 35 &  5.5 & &  &  & \star\star \\
6628 & 850 & 100 & 110 & 28.0 & 15.4 & 143 & 35 &  4.9 & & \star &  &  \\
6817 & 245 & 34 & 53 & 47.0 &  1.8 & 208 & 131 &  2.2 & &  &  &  \\
6944 & 3257 & 189 & 256 & 20.3 & 107.5 & 123 & 57 &  7.1 & & \star\star & \star\star &  \\
6956 & 917 & 96 & 122 & 14.0 &  8.2 & 140 & 38 &  3.3 & &  &  &  \\
7047 & 211 & 75 & 90 & 38.0 &  1.1 & 156 & 47 &  8.0 & & \star & \star\star &  \\
7125 & 1071 & 135 & 138 & 50.0 & 44.9 & 240 & 122 &  5.3 & & \star\star & \star\star & \star\star \\
7151 & 264 & 144 & 144 & 54.2 &  1.6 & 192 & 19 &  7.0 & &  &  & \star\star \\
7199 & 164 & 18 & 23 &  9.7 &  0.3 & 70 & 60 &  3.0 & &  &  &  \\
7232 & 230 & 55 & 80 & 24.5 &  0.7 & 112 & 68 &  7.8 & & \star &  &  \\
7261 & 853 & 153 & 138 & 34.3 &  6.7 & 159 & 35 &  5.2 & &  & \star\star & \star \\
7278 & 293 & 121 & 145 & 261.7 &  7.6 & 447 & 136 &  8.3 & &  &  &  \\
7323 & 518 & 149 & 157 & 47.8 &  7.4 & 184 & 35 &  4.1 & &  & \star\star &  \\
7399 & 535 & 194 & 218 & 44.5 &  7.4 & 192 & 82 &  7.9 & & \star\star & \star\star & \star \\
7408 & 462 & 19 & 29 &  9.1 &  1.5 & 85 & 61 &  2.0 & &  & \star\star &  \\
7490 & 467 & 161 & 166 & 17.5 &  3.0 & 142 & 44 &  3.3 & &  &  &  \\
7524 & 320 & 151 & 154 & 334.8 &  9.7 & 527 & 50 &  3.9 & & \star\star &  & \star\star \\
7559 & 218 & 70 & 75 & 30.3 &  0.7 & 156 & 60 &  3.6 & & \star\star & \star & \star\star \\
7577 & 196 & 32 & 43 & 28.3 &  0.8 & 165 & 75 &  2.1 & &  & \star\star &  \\
7592 & 202 & 191 & 248 & 689.2 & 19.9 & 698 & 147 & 10.0 & &  &  & \star \\
7603 & 641 & 126 & 129 & 49.1 &  5.4 & 192 & 74 &  6.3 & & \star & \star &  \\
7608 & 537 & 131 & 147 & 33.5 &  5.6 & 169 & 31 &  5.4 & & \star\star &  & \star \\
7690 & 537 & 114 & 126 & 24.8 &  3.7 & 140 & 46 &  8.8 & &  & \star &  \\
\noalign{\smallskip}\hline
\end{tabular}
\end{flushleft}
\end{table*}
\addtocounter{table}{-1}
\begin{table*}
{\vspace{-0.2cm}}
\caption[]{-- Continued}
\begin{flushleft}
\begin{tabular}{rcccccccccccc}
\noalign{\vspace{-0.2cm}}\hline\noalign{\smallskip}
 UGC & $\varv_\mathrm{sys}$ & $W_{R,20}^i$ & $W_{R,50}^i$ & $\int$Sdv & $M_\mathrm{HI}$ & $R_\mathrm{HI}$ & $h_\mathrm{HI}$ & $\langle\Sigma_\mathrm{HI}\rangle_{3.2h}$ & & \multicolumn{3}{c}{lopsidedness} \\
 & \kms & \kms & \kms & Jy \kms & $10^8$ \msun & $''$ & $''$ & M$_\odot$pc$^{-2}$ & & prof. & dens. & kin.\\
 (1) & (2) & (3) & (4) & (5) & (6) & (7) & (8) & (9) & & (10) & (11) & (12) \\
\noalign{\smallskip}\hline\noalign{\smallskip}
\omit\ruleit{1.2cm}&\ruleit{0.9cm}&\ruleit{0.9cm}&\ruleit{0.9cm}&\ruleit{1.05cm}&\ruleit{0.9cm}&\ruleit{0.9cm}&\ruleit{0.9cm}&\ruleit{0.9cm}&\ruleit{0.4cm}&\ruleit{0.9cm}&\ruleit{0.9cm}&\ruleit{0.9cm}\\
7866 & 359 & 62 & 71 & 23.6 &  1.3 & 149 & 34 &  4.7 & &  &  &  \\
7916 & 606 & 64 & 71 & 21.6 &  3.6 & 140 & 17 &  3.1 & &  &  &  \\
7971 & 467 & 84 & 99 & 16.3 &  2.7 & 107 & 27 &  5.7 & & \star &  &  \\
8188 & 313 & 87 & 116 & 57.6 &  3.0 & 193 & 38 &  5.4 & &  & \star\star &  \\
8201 & 37 & 43 & 58 & 35.0 &  2.0 & 175 & 72 &  3.3 & & \star &  &  \\
8286 & 407 & 165 & 163 & 65.0 &  3.5 & 256 & 62 &  4.3 & &  &  & \star \\
8331 & 260 & 47 & 55 & 17.3 &  1.4 & 179 & 35 &  1.7 & & \star & \star\star &  \\
8490 & 204 & 136 & 144 & 140.6 &  8.0 & 346 & 135 &  9.1 & & \star\star &  &  \\
8550 & 364 & 113 & 115 & 27.7 &  1.8 & 172 & 16 &  4.0 & &  &  &  \\
8683 & 658 & 47 & 49 &  8.5 &  3.2 & 94 & 26 &  3.2 & &  &  &  \\
8837 & 144 & 78 & 93 & 26.6 &  1.6 & 136 & 35 &  1.8 & & \star\star &  & \star\star \\
9128 & 154 & 43 & 54 & 13.9 &  0.6 & 87 & 37 &  7.5 & &  &  &  \\
9211 & 685 & 129 & 132 & 27.9 & 10.5 & 167 & 42 &  6.2 & &  &  &  \\
9992 & 427 & 73 & 79 & 11.8 &  3.0 & 105 & 29 &  5.3 & & \star\star &  & \star \\
10310 & 713 & 150 & 157 & 21.9 & 12.6 & 130 & 27 &  6.2 & &  &  & \star \\
11557 & 1389 & 155 & 174 & 18.9 & 25.2 & 142 & 41 &  5.0 & &  &  &  \\
11707 & 905 & 184 & 186 & 62.4 & 37.2 & 203 & 37 &  5.2 & & \star\star &  & \star\star \\
11861 & 1481 & 296 & 314 & 48.0 & 71.4 & 194 & 33 &  5.1 & &  &  &  \\
12060 & 884 & 159 & 171 & 31.0 & 18.1 & 161 & 58 &  4.0 & & \star\star & \star\star & \star \\
12632 & 422 & 141 & 145 & 77.1 &  8.7 & 266 & 43 &  3.4 & & \star\star &  & \star\star \\
12732 & 749 & 172 & 180 & 89.1 & 36.6 & 272 & 35 &  4.7 & & \star\star &  & \star\star \\
\noalign{\smallskip}\hline
 ~ & & & & & & & & & & & & \\
 ~ & & & & & & & & & & & & \\
 ~ & & & & & & & & & & & & \\
 ~ & & & & & & & & & & & & \\
 ~ & & & & & & & & & & & & \\
 ~ & & & & & & & & & & & & \\
 ~ & & & & & & & & & & & & \\
 ~ & & & & & & & & & & & & \\
 ~ & & & & & & & & & & & & \\
 ~ & & & & & & & & & & & & \\
 ~ & & & & & & & & & & & & \\
 ~ & & & & & & & & & & & & \\
 ~ & & & & & & & & & & & & \\
 ~ & & & & & & & & & & & & \\
 ~ & & & & & & & & & & & & \\
 ~ & & & & & & & & & & & & \\
 ~ & & & & & & & & & & & & \\
 ~ & & & & & & & & & & & & \\
 ~ & & & & & & & & & & & & \\
 ~ & & & & & & & & & & & & \\
 ~ & & & & & & & & & & & & \\
 ~ & & & & & & & & & & & & \\
 ~ & & & & & & & & & & & & \\
 ~ & & & & & & & & & & & & \\
 ~ & & & & & & & & & & & & \\
 ~ & & & & & & & & & & & & \\
 ~ & & & & & & & & & & & & \\
 ~ & & & & & & & & & & & & \\
 ~ & & & & & & & & & & & & \\
 ~ & & & & & & & & & & & & \\
 ~ & & & & & & & & & & & & \\
 ~ & & & & & & & & & & & & \\
 ~ & & & & & & & & & & & & \\
 ~ & & & & & & & & & & & & \\
 ~ & & & & & & & & & & & & \\
 ~ & & & & & & & & & & & & \\
\end{tabular}
\end{flushleft}
\end{table*}
}

\newpage\clearpage

\section{Atlas of the \HI\ observations}
\label{thefigures}

In the next pages overview figures are presented with the \HI\ data
for all the galaxies in the sample. For each galaxy a figure is given
with six panels. The \HI\ data shown are at $30''$ resolution.

\noindent {\it Top left.} A grayscale representation of the integrated
\HI\ distribution.  The contour levels are $1, 2, 5, 8, 11, 15, 20,
30, 40, 50,$ ... $10^{20}$ atoms cm$^{-2}$. The outermost contour
always represents a column density of $1\times 10^{20}$ atoms cm$^{-2}$.
Contours representing column densities of $5\cdot 10^{21}$ atoms
cm$^{-2}$ or higher are displayed in white.  The thick solid line
gives the approximate three sigma level, determined as described in
Sect.~\ref{thereduction}. The plus sign indicates the position of
the optical center.  The beam size is given in the lower left.

\noindent {\it Top right.} The velocity field. The thick line is the
systemic velocity. The interval between the thin lines is given in the
lower left. Light shading indicates receding velocities, dark shading
indicates approaching velocities. The extent of the velocity field may
be smaller than that of the integrated \HI\ map, because velocities
were only determined from profiles with a signal-to-noise ratio higher
than three.

%\vfill
%
%\hspace{-0.5cm}\hbox to\textwidth{
%\vtop{\hbox{\resizebox{0.5\textwidth}{!}{\includegraphics{h3074f13.ps}}}
%}\hfil
%\vtop{\hbox{\resizebox{0.5\textwidth}{!}{\includegraphics{h3074f14.ps}}}
%}}
%\vspace{0.4cm}
%
%\break

\noindent {\it Middle left.} The integrated \HI\ distribution overlayed
on the optical $R$-band image from Paper~II.

\noindent {\it Middle right.} The position-velocity diagram along the
major axis. Contour levels are at $-4$ and $-2$ (dotted),
and $2\sigma$, $4\sigma$, ... .

\noindent {\it Bottom left.} The radial \HI\ density profile. The thin
solid line gives the profile for the approaching side, the dotted line
gives the profile for the receding side. The thick solid line is the
average radial \HI\ density profile. The vertical arrow indicates a
radius of 3.2 optical disk scale lengths.

\noindent {\it Bottom right.}
The global profile. The arrow indicates the systemic velocity.

\vspace{1cm}

\noindent {\bf Full version with figures in this appendix can be
downloaded from:

\noindent http://www.robswork.net/publications/WHISPI.ps.gz}


\begin{thebibliography}{}

\bibitem[]{} Baldwin J. E., Lynden-Bell D., \& Sancisi R. 1980, MNRAS,
  193, 313

\bibitem[]{} Bosma, A. 1978, PhD thesis, Rijksuniversiteit Groningen

\bibitem[]{} Bosma, A. 1981a, AJ 86, 1791

\bibitem[]{} Bosma, A. 1981b, AJ 86, 1825

\bibitem[]{}
Broeils, A. H. 1992a, PhD thesis, Rijksuniversiteit Groningen

\bibitem[]{}
Broeils, A. H. 1992b, A\&A 256, 19

\bibitem[]{}
Broeils, A. H., \& Rhee, M.-H. 1997, A\&A 324, 877 (BR)

\bibitem[]{}
Broeils, A. H., \& van Woerden, H. 1994, A\&AS, 107, 129

\bibitem[]{}
Burstein, D., \& Heiles, C. 1984, ApJS, 54, 33

\bibitem[]{}
Carignan, C., \& Beaulieu, S. F. 1989, ApJ 347, 760

\bibitem[]{}
Carignan, C., \& Purton, C. 1998, ApJ 506, 125

\bibitem[]{}
Cayatte, V., van Gorkom, J. H., Balkowski, C., \& Kotanyi, C. G. 1990, AJ
100, 604

\bibitem[]{}
Cayatte, V., Kotanyi, C. G., Balkowski, C., \& van Gorkom, J. H. 1994, AJ
107, 1003 (CKBG)

\bibitem[]{}
C{\^o}t{\'e}, S., Carignan, C., \& Freeman, K. C. 2000, AJ 120, 3027

\bibitem[]{}
de Blok, W. J. G., McGaugh, S. S., \& van der Hulst, J. M. 1996, MNRAS 283,
18

\bibitem[]{} de Vaucouleurs, G., de Vaucouleurs, A., Corwin, H. G., et
  al. 1991, Third Reference Catalogue of Bright Galaxies (New
  York:Springer)

\bibitem[]{}
Freeman, K. C. 1970, ApJ 160, 811

\bibitem[]{}
Haynes M. P., Hogg D. E., Maddalena R. J., Roberts M. S., \& van Zee, L.
1998, AJ 115, 62

\bibitem[]{}
Hoffman, G. L., Salpeter, E. E., Farhat, et al. 1996, ApJS 105, 269

\bibitem[]{}
Hunter, D. A., Wilcots, E. M., van Woerden, H., Gallagher, J. S., \& Kohle,
S. 1998, ApJ 495, L47

\bibitem[]{}
Kraan-Korteweg, R. C. 1986, A\&AS 66, 255

\bibitem[]{}
Lucy, L. B. 1974, AJ 79, 745

\bibitem[]{}
Meurer, G. R., Carignan, C., Beaulieu, S. F., \& Freeman, K. C. 1996, AJ
111, 1551

\bibitem[]{}
Nilson, P. 1973, Uppsala General Catalogue of Galaxies, Uppsala
Astr. Obs. Ann., Vol. 6 (UGC)

\bibitem[]{}
Puche, D., Westpfahl, D., Brinks, E., \& Roy, J.R. 1992, AJ 103, 1841

\bibitem[]{}
Rhee, M.-H., van Albada, T.S. 1996, A\&AS 115, 407

\bibitem[]{}
Richter O.-G., Sancisi R. 1994, A\&A, 290, L9

\bibitem[]{}
Rieke, G. H., \& Lebofksy, M. J. 1985, ApJ, 288, 618

\bibitem[]{}
Roberts, M. S. 1969, M.S. AJ 74, 859

\bibitem[]{}
Roberts, M. S., \& Haynes, M. P. 1994, ARA\&A 32, 115

\bibitem[]{}
Swaters, R. A. 1999, PhD thesis, Rijksuniversiteit Groningen

\bibitem[]{}
Swaters, R. A., \& Balcells, M. 2000, A\&AS, accepted (Paper~II)

\bibitem[]{}
Swaters, R. A., Schoenmakers, R. H. M., Sancisi, R., \& van Albada, T. S.
1999, MNRAS 304, 330

\bibitem[]{}
Tully, R. B., \&  Fouqu\'e, P. 1985, ApJS., 58, 67

\bibitem[]{}
Verheijen, M. A. W., \& Sancisi, R., 2001, A\&A, 370, 765

\bibitem[]{}
Warmels, R. H. 1988a, A\&AS 72, 19

\bibitem[]{}
Warmels, R. H. 1988b, A\&AS 72, 427

\bibitem[]{}
Warmels, R. H. 1988c, A\&AS 73, 453

\bibitem[]{}
Wevers, B. M. H. R. 1984, PhD thesis, Rijksuniversiteit Groningen

\end{thebibliography}
\end{document}